\documentclass[aps,pra,twocolumn,showpacs,longbibliography,superscriptaddress]{revtex4-1}
\usepackage{titlesec}
\usepackage{amsmath}
\usepackage{amssymb}
\usepackage[fleqn]{mathtools}
\usepackage{braket}
\usepackage{todonotes}
\usepackage{bm}
\usepackage{braket}
\usepackage{color}
\usepackage{graphicx}
\usepackage{epstopdf}
\usepackage{xcolor}

\usepackage[labelformat=simple]{subcaption}

\titlespacing\section{0pt}{12pt plus 4pt minus 2pt}{12pt plus 4pt minus 2pt}
\titlespacing\subsection{0pt}{12pt plus 2pt minus 2pt}{6pt plus 2pt minus 2pt}
\begin{document}
\title{Quantum tetrachotomous states:\\
	Superposition of four coherent states on a line in phase space}
\author{Namrata Shukla}
\email{namrata.shukla1@ucalgary.ca}
\affiliation{Institute for Quantum Science and Technology, University of Calgary, Calgary, Alberta, T2N 1N4, Canada}
\author{Naeem Akhtar}
\email{naeemakhatr@mail.ustc.edu.cn}
\affiliation{%
	Shanghai Branch, National Research Center for Physical Sciences at Microscale,
	University of Science and Technology of China, Shanghai 201315, China%
    }
\author{Barry C. Sanders}
\email{sandersb@ucalgary.ca}
\affiliation{%
	Institute for Quantum Science and Technology,
	University of Calgary, Calgary, Alberta, T2N 1N4, Canada}
\affiliation{Program in Quantum Information Science, Canadian Institute for Advanced Research, Toronto, Ontario M5G 1M1, Canada}
\date{\today}
\begin{abstract}
The well studied quantum optical Schr\"{o}dinger cat state is a superposition of two distinguishable states,
with quantum coherence between these macroscopically distinguishable states being of foundational and,
in the context of quantum-information processing,
practical use.
We refer to these quantum-optical cat states as quantum dichotomous states,
reflecting that the state is a superposition of two options,
and we introduce the term quantum multichotomous state to refer to a superposition of multiple macroscopically distinguishable options.
For a single degree of freedom, such as position,
we construct the quantum multichotomous states as a superposition of Gaussian states
on the position line in phase space.
Using this nomenclature,
a quantum tetrachotomous state (QTS)
is a coherent superposition of four macroscopically distinguishable states.
We define, analyze and show how to create such states,
and our focus on the QTS is due to their exhibition of much richer phenomena than for the quantum dichotomous states with lessons to going to general multichotomous states with the quantum comb state as a limiting case.
Our characterization
of the QTS involves the Wigner function,
its marginal distributions, 
and the photon-number distribution,
and we discuss the QTS's approximate realization in a multiple coupled-well system.
\end{abstract}
\maketitle
\section{Introduction}
\label{sec:intro}
Schr{\"o}dinger's original quantum-cat paradox arose from the puzzle that quantum mechanics has enormous ramifications on dichotomous states such as life and death for a cat~\cite{Sch26},
and this notion morphed into the simplified problem of a superposition of two macroscopically distinguishable (i.e., negligibly overlapping) coherent states,
each representing a classical distinct choice.
Schr\"{o}dinger's coherent states are essentially non-spreading wave-packet solutions of the quadratic harmonic-oscillator potential~\cite{Sch35}.
Milburn first showed the $Q$-function (Husimi distribution~\cite{Hus40})
for this so-called cat state but did not discuss it explicitly~\cite{Mil86}.

Yurke and Stoler explicitly studied this fascinating superposition state~\cite{YS86}
via the position distribution
\begin{equation}
\label{eq:prq}
	\operatorname{pr}(q)
		=\left|\left\langle q|\psi\right\rangle\right|,\;
	\psi(q)\in L^2(\mathbb{R}),\;
	q\in\mathbb{R}
\end{equation}
and the momentum distribution
\begin{equation}
\label{eq:prp}
	\widetilde{\operatorname{pr}}(p)
		=\left|\left\langle q|\tilde{\psi}\right\rangle\right|,\;
	\tilde{\psi}(p)=\int_\mathbb{R}\text{d}q\,\text{e}^{\text{i}qp},\;
	p\in\mathbb{R},\;
	\hbar\equiv1.
\end{equation}
These canonical-variable distributions~(\ref{eq:prq}) and~(\ref{eq:prp})
are obtained from the Wigner function~\cite{Wig32,MOSW84,Sch01,Leo13}
\begin{equation}
\label{eq:Wignerqp}
	W(q, p)
		=\frac{1}{2\pi}\int_\mathbb{R}\text{d}x~\text{e}^{\text{i}px}
			\psi^*\left(q+\frac{x}{2}\right)\psi\left(q-\frac{x}{2}\right),
\end{equation}
as marginal distributions~\cite{Leo13}
\begin{equation}
\label{eq:Wmarginal distributions}
	\operatorname{pr}(q)
		=\int_\mathbb{R}\text{d}p~W(q, p),\;
	\widetilde{\operatorname{pr}}(p)
		=\int_\mathbb{R}\text{d}q~W(q, p),
\end{equation}
with these properties~(\ref{eq:Wmarginal distributions})
being strong motivation for representing states by Wigner functions.
Wigner functions have been used to study superpositions of Gaussian states~\cite{BK91,BW98}.

The study by Milburn~\cite{Mil86},
elaborated and extended by Milburn and Holmes~\cite{MH86},
and studied in the context of quantum-optical cat states by Yurke and Stoler~\cite{YS86}
all focused on superpositions of Gaussian states
\begin{equation}
\label{eq:gaussianstate}
	\vartheta\left(x;\mu,\sigma\right)
		=\sqrt{G\left(x;\mu,\sigma\right)}
\end{equation}
for
\begin{equation}
\label{eq:gaussian}
	G\left(x;\mu,\sigma\right)
		\coloneqq \frac{1}{\sqrt{2\pi}\sigma}
			\exp\left\{-\frac12{\Big(\frac{x-\mu}{\sigma}\Big)}^2\right\},
\end{equation}
with position distribution
\begin{equation}
\label{eq:gaussianstateposdist}
	\left|\vartheta\left(x;\mu,\sigma\right)\right|^2
		=G\left(x;\mu,\sigma\right).
\end{equation}
In the special case of coherent states~\cite{Gla63},
\begin{equation}
\label{eq:coherentstate}
	\left\{\ket{\alpha};\alpha\in\mathbb{C},
		\alpha(x)=\vartheta\left(x;\mu,\sqrt2\right)\right\}.
\end{equation}

In the Fock-state basis~$\{\ket{n};n\in\mathbb{N}\}$,
the coherent state is
\begin{equation}
\label{eq:coherentstateFockbasis}
	\ket{\alpha}
		=\text{e}^{-\frac{{\left|\alpha\right|}^2}{2}}\sum_{n=0}^\infty
			\frac{\alpha^n}{\sqrt{n!}}\ket{n},
\end{equation}
leading to the Poisson-number distribution
\begin{equation}
\label{eq:pndcs}
	\wp_\text{cs}(n;\alpha)
		:=\text{e}^{-\left|\alpha\right|^2}
			\frac{\left|\alpha\right|^{2n}}{n!}
\end{equation}
with mean
\begin{equation}
\label{eq:<n>}
	\langle n\rangle=\left|\alpha\right|^2
\end{equation}
and variance of
\begin{equation}
\label{eq:Vn}
	\left\langle\left(n-\langle n\rangle\right)^2\right\rangle=\left|\alpha\right|^2
\end{equation}
equal to each other both.

In quantum optics,
the Schr\"{o}dinger cat state typically refers to the (unnormalized) superposition of coherent states
\begin{equation}
\label{eq:cat}
	\ket{\alpha}+\text{e}^{\text{i}\varphi}\ket{-\alpha}.
\end{equation}
The cases of $\varphi=0$ and $\varphi=\pi$ are known as the even and odd coherent states, respectively~\cite{DMM74},
and $\varphi=\pi/2$ corresponds to the case that the cat state has a Poissonian photon-number distribution~\cite{YS86}.

Variants have been studied such as superpositions of squeezed states~\cite{San89},
which are essentially the full gamut of Gaussian states~(\ref{eq:gaussianstate}),
phase states~\cite{San92Jun},
significantly overlapping coherent states
(hence only partially distinguishable and known as kitten states)~\cite{OTL+06,OJT+07},
multimode coherent states also known as entangled coherent states~\cite{San92May,San92E,San12,WGR+16},
and superpositions generalizing coherent states
from rank-one SU(2) and SU(1,1) symmetries~\cite{GG97}
to higher-rank groups such as SU(3)~\cite{NS01}.

Almost all studies of coherent-state superpositions stay true to the essence of Schr\"{o}dinger's life-death dichotomy,
but a few studies ventured into quantum multichotomous states.
Superpositions of coherent states on the circle~\cite{BVK92, MTK90, JDA93},
with applications such as showing the limit yields a coherent-state representation of the Fock number state on the circle~\cite{Gar85},
form one branch of studies of superpositions of multiple coherent states.

Another investigative branch of superpositions of multiple coherent states 
corresponds to the superposition of Gaussian states,
e.g., superpositions of coherent states and and superpositions of squeezed states,
on the real line in phase space~\cite{GKP01}.
Such states can be studied as an unnormalizable superposition of an infinite number of equally spaced coherent states represented in phase space,
which we refer to as a ``comb state''~\cite{GKP01},
and also studied under realistic limitations~\cite{GS07}.
A particular case of superpositions of coherent states on the line in phase space,
which explicitly shows what we call a quantum tetrachotomous state (QTS),
which are superpositions of four negligibly overlapping quantum states,
as well as superpositions of a continuum of coherent states~(\ref{eq:coherentstate}),
were studied for the squeezing properties~\cite{BVK92}.

Here we study QTSs as special cases of quantum multichotomous states yet capturing the rich phenomena present in such states but being sufficiently simple to capture the richness of such states.
Henceforth,
we write the QTS as the state $\ket{\Upsilon}$
and its position representation as
\begin{equation}
\label{eq:Upsilonx}
	\Upsilon(x)
		=\langle x|\Upsilon\rangle.
\end{equation}
As we expect that the QTS features will be sensitive to separation of coherent-state amplitudes in phase space,
we focus on three exemplary cases of the QTSs.

These three cases are symmetric about $x=0$;
i.e.,
\begin{equation}
\label{eq:Upsilon}
	\Upsilon(x)=\Upsilon(-x).
\end{equation}
This symmetry simplifies the expressions and readily displays the intriguing features.
Furthermore, we restrict the QTS to a superposition of coherent states~(\ref{eq:coherentstate})
as this simplification is common, but not universal,
for quantum-optical cat states.

We treat three cases of the QTSs as a superposition of two doublets,
where we use the term doublet to refer to a superposition of a pair of coherent states.
\begin{itemize}
\item[$\Upsilon$1]
\label{case:QTS1}
	The first QTS case is a superposition of two doublets;
i.e., two coherent states whose amplitudes are close together in phase space
(kitten state~\cite{OTL+06,OJT+07})
whereas each of the two (kitten) doublets is macroscopically distinguishable from the other (kitten) doublet.
\item[$\Upsilon$2]
\label{case:QTS2}
	In the second case,
	each of the two doublets is macroscopically distinguishable,
but the pair of doublets are close in phase space,
i.e., only microscopically distinguishable.
\item[$\Upsilon$3]
\label{case:QTS3}
	The third case corresponds to equal separation between coherent states on the line, which looks like a four-peaked comb state in the phase-space representation.
\end{itemize}

Our approach to characterizing QTSs employs Wigner functions~\cite{BK91},
canonical position measurements~\cite{VR89}
and photon-number distributions~\cite{Sch01},
given by
\begin{equation}
\label{eq:Pin}
	\wp_\text{QTS}(n;\alpha,\beta)=\left|\langle n|\Upsilon\rangle\right|^2
\end{equation}
for a QTS with~$\ket{n}$ the Fock state with~$n$ photons.
We employ the Wigner function to analyze QTSs because
the Wigner function gives meaning to talking about states as being ``in phase space'',
i.e., the phase-space representation.
Essentially,
the QTS~$\ket{\Upsilon}$ should appear,
in the Wigner-function representation,
as a sum of displaced Gaussian distributions plus additional features that reveal quantum effects such as interference due to coherence between Gaussian states.

Notably,
negativity of the Wigner function signifies quantum effects~\cite{DGBR15},
and marginal distributions of the Wigner function
directly yield canonical position distribution arising as asymptotic limits of homodyne detection 
for strong local-oscillator fields~\cite{SYM79,YS80,TS04}.
The photon-number distribution
is interesting as phase-space interference can lead to oscillations in the photon-number distribution~\cite{SW87},
which has been studied for general quantum-optical cat states~\cite{BVK92, BGK93, SPK91}
including for entangled coherent states~\cite{AM94,DMN95}.

Various concepts for realizing quantum-optical cat states can be extended to TQSs.
Caldeira and Leggett studied tunnelling in a dissipative system with superconductors
as the medium~\cite{CL83}, 
and a cat-like state can be achieved at half the tunnelling time between the two wells.
Essentially, a cat-like state is approximated by the 
ground state of a double-well potential.
A QTS could arise as the ground state of a multiple coupled-well potential,
as we show in this work.
One way to realize the ground state of a multiple coupled-well potential can be considered as ground state of coupled Bose-Einstein condensates as a generalization of Bose-Einstein condensates in a double-well potential~\cite{SC98, CLMZ98, Ruo01}.

Our paper is organized as follows. In~\S\ref{sec:background}, we present background for quantum-optical cat states
including characterization using Wigner functions, canonical marginal distributions
and photon-number distributions. 
Our background section also includes the realization of quantum-optical cat states as approximate ground state of a double-well potentials plus experimental generation of quantum-optical cat states. In~\S\ref{sec:approach}, we describe our approach to analyzing the QTS.
Analysis and results appear in~\S\ref{sec:analysis and results}, \S\ref{sec:discussion} contains the discussion and \S\ref{sec:conclusions} is for conclusions.
\section{Background}
\label{sec:background}
In this section, we discuss the theory of quantum-optical cat states,
characterizations of such states,
and their experimental creation.
Furthermore, 
we present the analogy between quantum-optical cat states with the ground state of a double-well potential.
\subsection{Theory}
\label{subsec:theory}
In this subsection, we define the quantum-optical cat state as a superposition of two coherent states~(\ref{eq:cat}) with emphasis on even and odd quantum-optical cat states.
We present characterization  of quantum-optical cat states including the Wigner function and the photon-number distribution,
and discuss theoretical proposals for generating quantum-optical cat states.
Additionally,
we discuss the marginal distribution,
the inverse radon transform and tomography.
Finally, 
we discuss the approximate realization of this state as the ground state of a double-well potential.

The quantum-optical cat state has been characterized using various quasi-probability distributions~\cite{Gla63, Sud63, Hus40, CG69}
that are advantageous because they can be compared to classical systems represented in phase space.
The Wigner function is now a typical tool for studying quantum-optical cat states and their interference pattern~\cite {BVK92, BW98}.
Cat states are typically characterized by performing rotated position quadrature measurements~\cite{VR89} or,
in quantum optics language,
homodyne measurements from which the Wigner function can be constructed via, e.g., by the inverse radon transform~\cite{Leo13}.

\begin{figure}
\begin{subfigure}{0.23\textwidth}
\includegraphics[width=\linewidth]{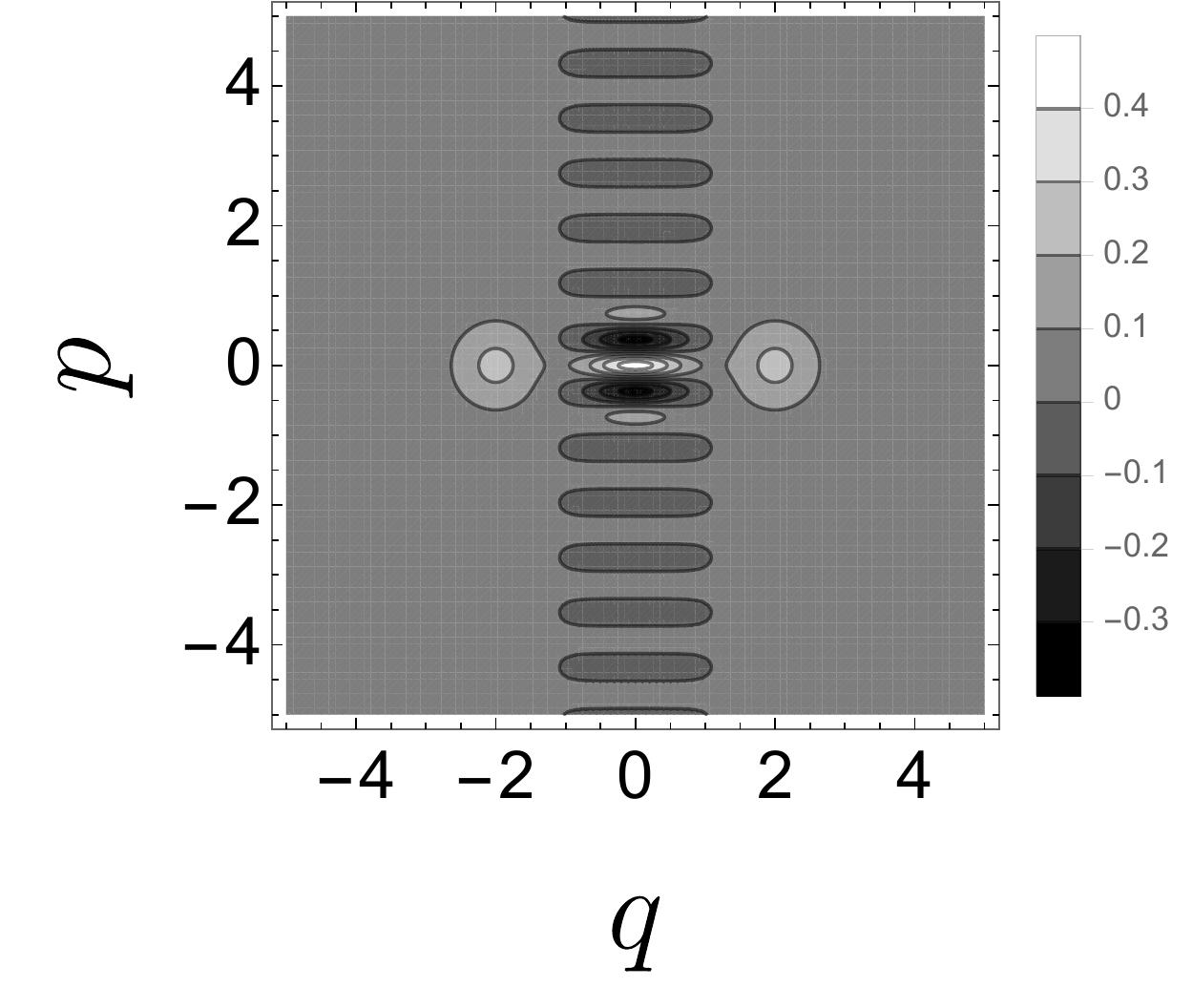} 
\caption{ }
\label{subfig:Wcat1}
\end{subfigure}
\hspace{0 cm}
\begin{subfigure}{0.23\textwidth}
\includegraphics[width=\linewidth]{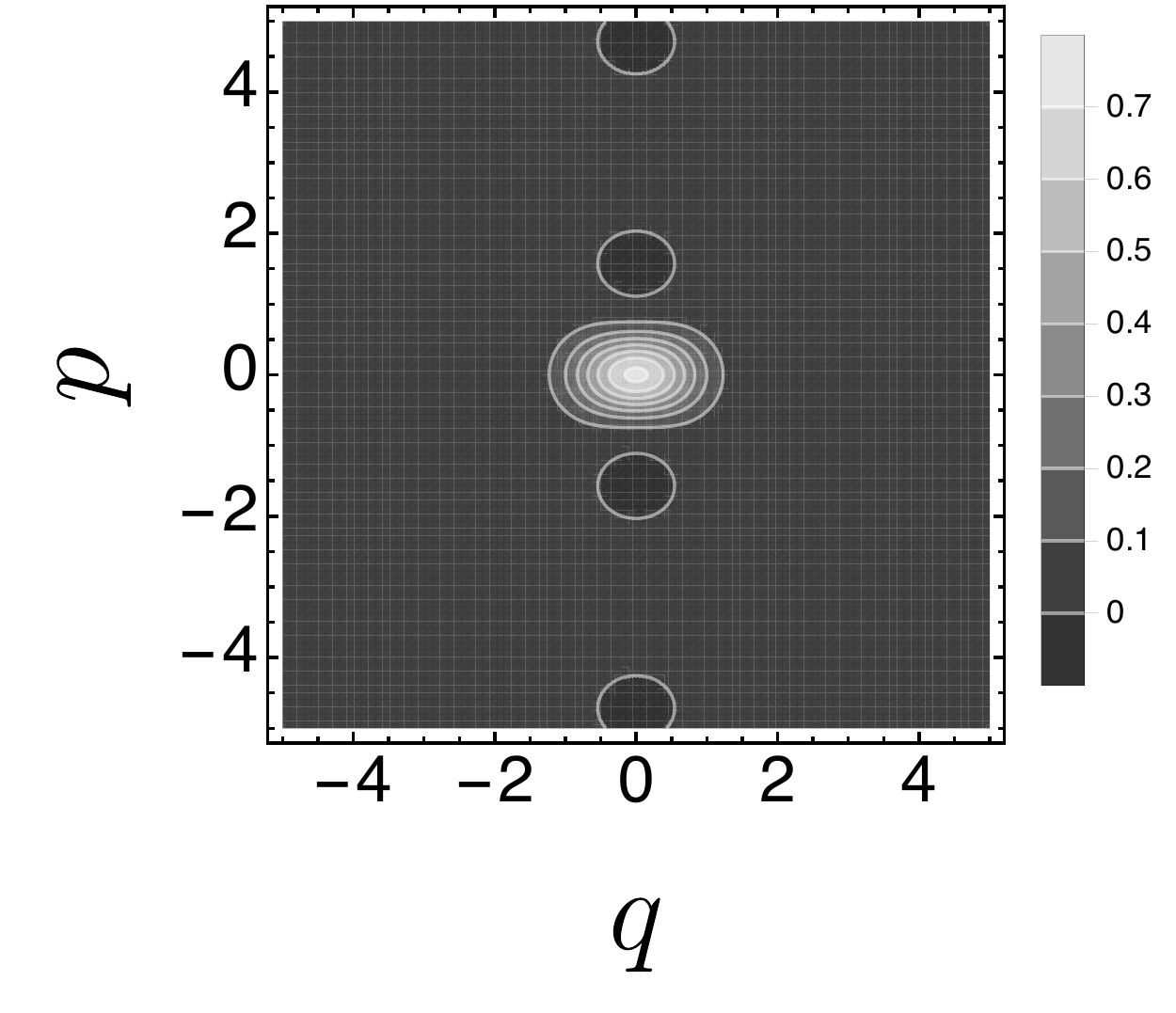}
\caption{ }
\label{subfig:Wcat2}
\end{subfigure}
\caption{Wigner function for $\ket{\alpha}+\ket{-\alpha}$
with~\subref{subfig:Wcat1}
$\alpha=2$ (cat state)
and~\subref{subfig:Wcat2} $\alpha=0.5$ (kitten state).%
}
\label{fig:Wcat}
\end{figure}

Depending on the separation between coherent states~$\ket{\pm\alpha}$,
which is given by the distance~$2\left|\alpha\right|$ in phase space,
their superpositions are referred as a quantum-optical cat state or as a kitten state
for large or small inter-state separation compared to the width of the coherent-state. Wigner function for this superposition is 
\begin{align}
\label{eq:cohstW}
	W_\text{cs}(q,p;\alpha)
		=&\frac{1}{2\pi}\int_\mathbb{R}\text{d}x~\text{e}^{\text{i}px}
			\left\langle q-\frac{x}{2}\Big|\alpha\right\rangle
				\nonumber\\
		=&\frac{1}{\sqrt{2}\pi}\text{e}^{-\frac{p^2}{2}-2(q+\alpha)^2}
				\\
		=&\frac{1}{\sqrt{2}}
			G\left(p;0,1\right)G\left(q;-\alpha,\frac12\right)
				\nonumber
\end{align}
with Gaussian distribution $G$~(\ref{eq:gaussian}).
Contours of the Wigner functions for even and odd quantum-optical cat states
are shown in Fig.~\ref{fig:Wcat}.
Yurke and Stoler~\cite{YS86} used the marginal distributions as canonical phase-space-rotated position and momentum distributions when they first explained the quantum-optical cat states. 

The Wigner function for a quantum-optical cat state is
\begin{align}
\label{eq:Wcat}
	&W_\text{cat}(q,p;\alpha)\nonumber\\
		=&\frac{1}{2\pi N_\alpha} \int_\mathbb{R}\text{d}x~\text{e}^{\text{i}px}
			\left[\left\langle q-\frac{x}{2}\Big|\alpha\right\rangle
				+\left\langle q-\frac{x}{2}\Big|-\alpha\right\rangle\right]
					\nonumber\\
		&\times\left[\left\langle\alpha\Big|q+\frac{x}{2}\right\rangle
			+\left\langle-\alpha\Big|q+\frac{x}{2}\right\rangle\right],
\end{align}
which is simplified to
\begin{align}
\label{eq:Wcateqsimp}
	W_\text{cat}(q, p;\alpha)
		=&\frac{\text{e}^{-\frac{p^2}{2}}}{\sqrt{2}\pi N_\alpha}
			\bigg[\text{e}^{-2(q+\alpha)^2}
			+\text{e}^{-2\left(q-\alpha\right)^2}
					\nonumber\\
		&+\underbrace{\text{e}^{-2q^2-2\text{i}p\alpha}
		+\text{e}^{-2q^2+2\text{i}p\alpha}}_{2\text{e}^{-2q^2}\cos2p\alpha}\bigg],
\end{align}
for normalization $N_\alpha=2(1+\text{e}^{-2\alpha^2})$ for real-valued even coherent state $\ket{\alpha}+\ket{-\alpha}$.
Using the Gaussian-distribution notation~(\ref{eq:gaussian}),
and introducing
\begin{equation}
	G_+\left(x;\gamma,\sigma\right)
		\coloneqq G\left(x;\gamma,\sigma\right)+G\left(x;-\gamma,\sigma\right),
\end{equation}
the Wigner function for the quantum-optical cat state is
\begin{align}
\label{eq:Wcatgaussian}
	W_\text{cat}(q,p;\alpha)
		=&\frac{1}{\sqrt{2} N_\alpha}
			\bigg[G\left(p;0,1\right)G_+\left(q;\alpha,\frac12\right) \nonumber\\
		&+\text{e}^{-2\alpha^2}G\left(q;\alpha,\frac12\right)G_+\left(p;2\text{i}\alpha,1\right)\bigg]. 
\end{align}
The first two terms in Eq.~(\ref{eq:Wcatgaussian})
are the Gaussian distributions corresponding to the coherent states,
and the last two terms are responsible for the interference pattern
between these two coherent states in the phase space.
The interference pattern can be explained by the periodic oscillatory functions appearing as coefficients to the Gaussians.

\subsection{Quantum-state tomography and marginal distributions}
\label{subsec:tomography}
For quantum-state tomography,
marginal distributions for the Wigner function can be used to estimate the Wigner function via the inverse radon transformation~\cite{Leo13}.
Operationally,
a series of homodyne measurements,
obtained for many choices of local-oscillator phase,
yields data that are then inserted into an algorithm that calculates an approximation to the state's Wigner function.

The position and momentum marginal distributions for the quantum-optical cat state are
\begin{equation}
\label{eq:prqcat}
	\operatorname{pr}_\text{cat}(q)
		=\frac{1}{N_\alpha}G_+^2\left(q;\alpha,\frac{1}{\sqrt{2}}\right),
\end{equation}
and
\begin{equation}
\label{eq:prpcat}
	\widetilde{\operatorname{pr}}_\text{cat}(p)
		=\frac{2\sqrt{2}}{N_\alpha}G(p;0,1)\cos^2{p\alpha},
\end{equation}
which are a mixture of two Gaussians and a sinusoidally modulated Gaussian,
respectively.
\begin{figure}
\begin{subfigure}{0.26\textwidth}
\includegraphics[width=\linewidth]{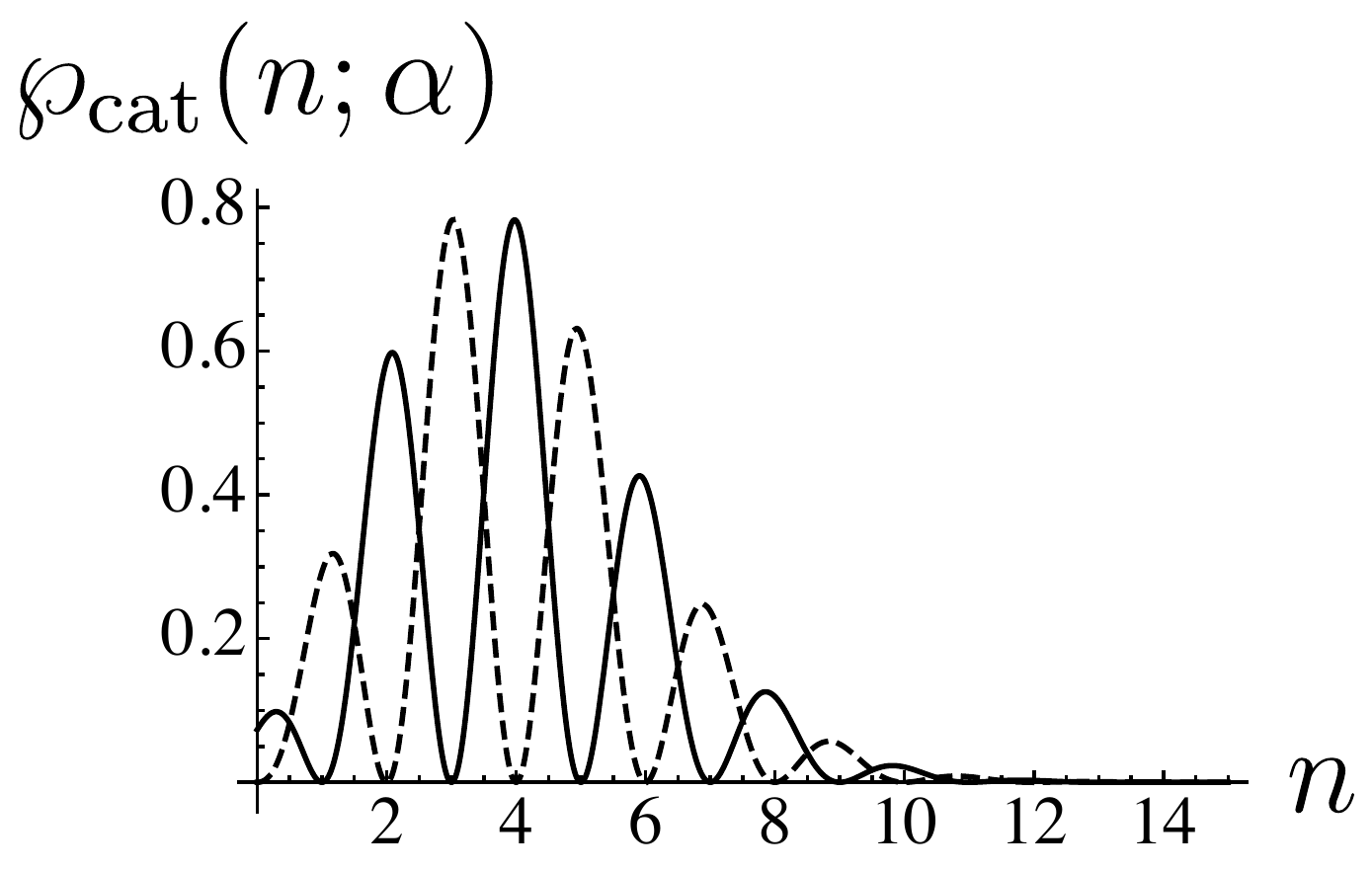} 
\caption{ }
\label{subfig:pndevenoddcoh1}
\end{subfigure}
\hspace{0 cm}
\begin{subfigure}{0.20\textwidth}
\includegraphics[height=2.87cm,width=\linewidth]{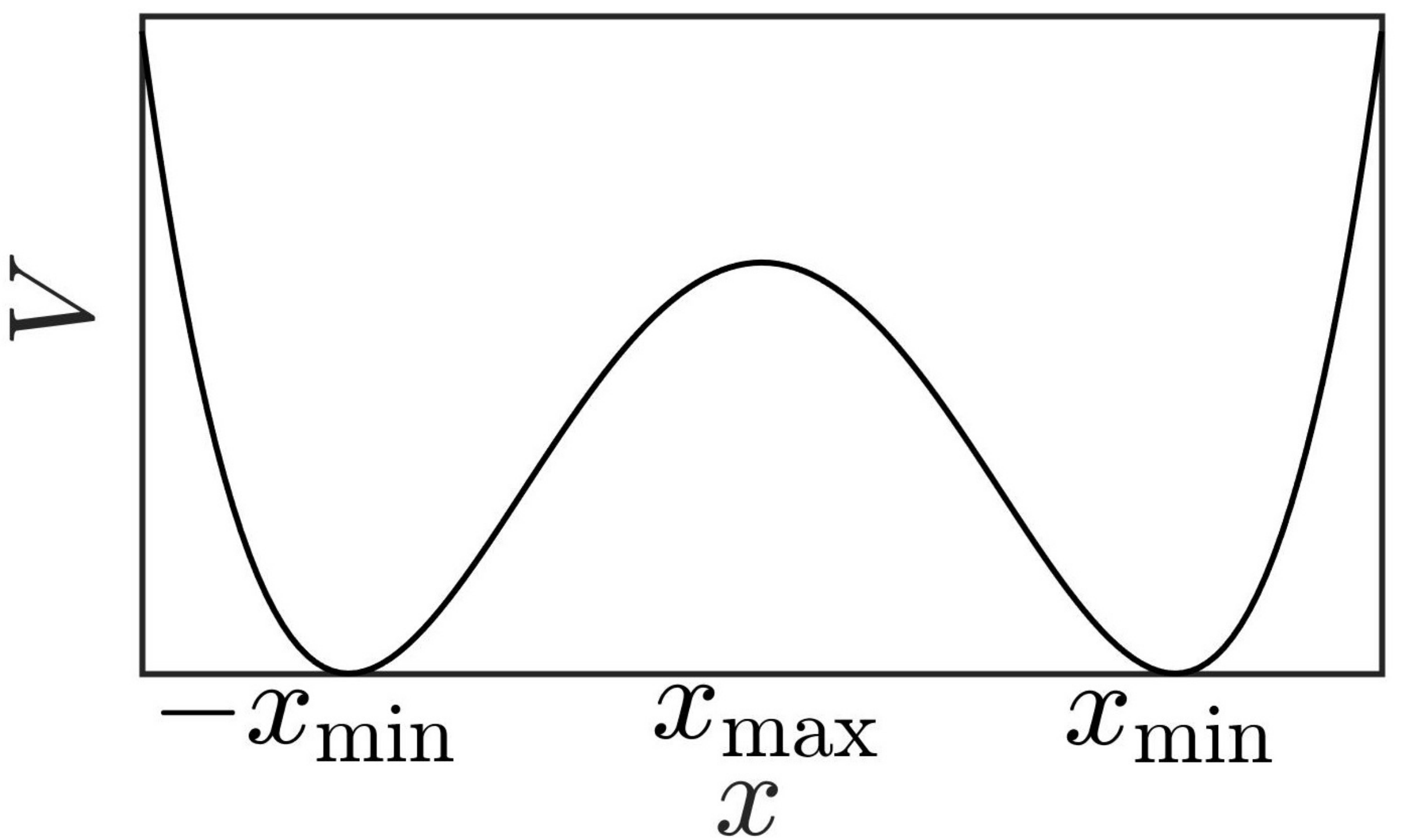}
\caption{ }
\label{subfig:pndevenoddcoh2}
\end{subfigure}
\caption{%
~\subref{subfig:pndevenoddcoh1} Photon-number distribution ($\alpha=2$) for even coherent state $\ket{\alpha}+\ket{-\alpha}$ (solid line)
and odd coherent state $\ket{\alpha}-\ket{-\alpha}$ (dashed line).~\subref{subfig:pndevenoddcoh2} Potential energy as a function of position $x$ with local minima $\pm x_\text{min}$ and local maximum $x_\text{max}$.}
\label{fig:pndevenoddcoh}
\end{figure}
\subsection{Interference in phase space}
\label{subsec:interference_phase space}
The photon-number distribution for a general superposition of coherent states reveals interference in phase space~\cite{BVK92, BGK93}.
This interference effect arises from the term that is sensitive to the relative phase difference between superposed coherent states.
The photon-number distribution for an even ($+$) and odd ($-$) cat state~(\ref{eq:cat})
is
\begin{equation}
\label{eq:pndcsgaussian}
	\wp_\text{cat}(n;\alpha)
		\propto
			\left[1\pm(-1)^n\right]
				\wp_\text{cs}(n;\alpha),
\end{equation}
with~$\wp$ the Poisson distribution~(\ref{eq:pndcs}).
This Poisson distribution is shown in Fig.~\ref{fig:pndevenoddcoh}.
Due to interference,
even coherent states exhibit only an even number of photons.
For the odd coherent states,
we observe only odd numbers of photons~\cite{DMN95,SW87}.
The photon-number distribution is modulated by interference between the two coherent states in the superposition.

Oscillations in the photon-number distribution are due to interference in phase space~\cite{SW87,Sch01},
and this kind of oscillation~(\ref{eq:pndcsgaussian})
has been studied
for superpositions of coherent states on a line and superposition of coherent states on a circle in phase space~\cite{JDA93}.
A closer the coherent states to each other, the less distinguishable the peaks
are from the interference fringes as seen in Fig.~\ref{subfig:Wcat2}.
As the separation between the Gaussian peaks on the phase-space position line
decrease to the point that the interferences fringes and peaks lose distinguishability,
momentum squeezing becomes evident.
\subsection{Generating quantum-optical cat states}
\label{subsec:qocatstates}
The original proposals for generating optical cat states~(\ref{eq:cat})
were achieved via nonlinear interactions~\cite{GNM+04, MT87, BHR92, LLNJ12}
subsequently followed by proposals for manipulating photons in a cavity by dispersive atom-field coupling~\cite{SBW90}.
Generating quantum-optical cat states by photon subtraction
is an appealing approach~\cite{LJRK04, DAO+97}.
An approximation to quantum-optical Schr{\"o}dinger kitten states can be prepared with a squeezed-light resource~\cite{OTL+06},
and a squeezed Schr{\"o}dinger cat state
has been approximately created
using homodyne detection and photon number states as resources~\cite{OJT+07}.

Cat states could be created by two interacting Bose condensates~\cite{CLMZ98}.
Alternatively,
cat states could be realized with two trapped, coupled Bose-Einstein condensates with a Josephson coupling~\cite{SC98}.
One enticing method to create a quantum-optical cat state,
i.e., a quantum dichotomous state (QDS),
is via constructing a ground state of a coherently coupled Bose-Einstein condensate in a double-well potential by means of scattering light and quantum measurement~\cite{Ruo01}.
We discuss the double-well potential next.
\subsection{Double-well potential}
\label{subsec:doublewell}
\label{subsec:doublewellpotential}
In this subsection,
we discuss the coherent state as the ground state of a quadratic potential and its extension to a QDS as the ground state of a double-well potential.
Quantum tunnelling is valuable for many applications,
with one foundational example being the case studied
by Caldeira and Leggett
to ascertain how long range quantum coherence can be~\cite{CL83,Leg80}.

A superposition of coherent states~(\ref{eq:cat})
can be created for large amplitudes $\left|\alpha\right|\gg1$
using the motional state of trapped ions in an approximate harmonic-oscillator potential
using ultrafast laser pulses~\cite{JWN+17} and coupling a qubit with a harmonic oscillator~\cite{LHT16}.
This superposition of coherent states can also be constructed as the ground state of a coherently coupled Bose-Einstein condensate in a double-well potential~\cite{SC98, CLMZ98} or the evolution of a Bose-Einstein condensate in a double-well potential in a two-mode approximation~\cite{Ruo01}. In the latter case, macroscopic quantum coherence of Bose-Einstein condensates leads to coherent quantum tunneling of atoms between the two modes representing two Bose-Einstein condensates, thereby creating a QDS.

Other experimental proposals include conditional generation of QDSs~\cite{STT08, GGC+10, NNH+06, ANS+17,TWS+08, NNH+06, STT08, GGC+10, EBK+15, ANS+17,OFT+09,EBK+15,TWS+08}
with applicability to quantum-information processing~\cite{VKL+13, DMH+17,Har13}.
Creation of QDSs, with the crucial macroscopic separation required to observe the quantum effects, is thus based on some degree of approximation such as the method proposed by Caldeira and Leggett~\cite{CL83}.

Schr\"{o}dinger showed that coherent states follow the motion of a classical particle in a quadratic potential~\cite{Sch26,Sch68},
and the ground state is a coherent-state with zero mean amplitude.
For a double-well system,
coupling two nearly harmonic oscillators together via a finite barrier,
which is approximated by an inverted harmonic oscillator in the vicinity of the unstable maximum potential between the two wells,
the localized ground state in each well can be approximated by a coherent state~(\ref{eq:coherentstate}),
with mean amplitude corresponding to the displacement of the well from the origin of phase space.
The harmonic-oscillator assumptions for the minima and maximum (inverted harmonic potential in that case)
are quite good for analytic potentials due to convergence of the Taylor series around minima and maxima.
The minima are stable points with zero-motion solutions classically and the maximum is a turning point classically.
In the rest of the paper we refer to stable and unstable points in phase space and their respective localized states in those regions.

Nandi \cite{Nan11} has proposed a way of creating a double well potential by multiplying the Gaussian~(\ref{eq:gaussian}) by a factor of $x^2$.
This new potential function, which is essentially a second-order Hermite-Gaussian function with a pure Gaussian subtracted,
is symmetric about $x=0$.
In this way,
the Hamiltonian can be brought into a symmetric tridiagonal form.
This system has been numerically solved to study the energy levels. The tunnelling rate of the particle in such a potential well depends on the energy difference between the ground and first excited state. The higher or wider the barrier the smaller this energy difference and this difference will become larger as the energy of the incident particle increases.

Equipped with this background and the limitations to construct and characterize Schr{\"o}dinger cat state (QDS), we now develop our concept of the QTS. The QTS as we define in the next section is in some sense a superposition of two QDSs. In the next section, we also discuss the tools and methods to analyze QTS and study their appearance and behaviour in the phase space.

\section{Approach}
\label{sec:approach}
In this section, we describe our approach to analyzing QTSs.
Our approach involves characterization by the Wigner function,
the marginal distribution and the approximate realization of the QTS states via a multiple coupled-well system, i.e., quadrupole-Gaussian-well potential. We also study the photon-number distribution of the QTS state and observe the oscillatory behaviour due to the four level interference patterns between the coherent states.

\subsection{Defining the tetrachotomous cat}
\label{subsec:definesub}
\begin{figure}
\centering
\begin{subfigure}{0.40\textwidth}
\includegraphics[width=\linewidth]{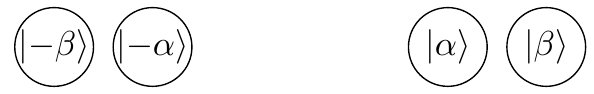} 
\caption{Case~1}
\label{subfig:phasespace picture1}
\vspace{0.5 cm}
\end{subfigure}
\begin{subfigure}{0.40\textwidth}
\includegraphics[width=\linewidth]{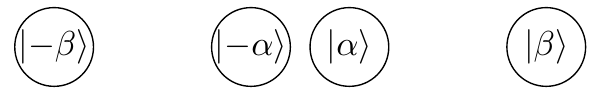}
\caption{Case~2}
\label{subfig:phasespace picture2}
\vspace{0.5 cm}
\end{subfigure}
\begin{subfigure}{0.40\textwidth}
\includegraphics[width=\linewidth]{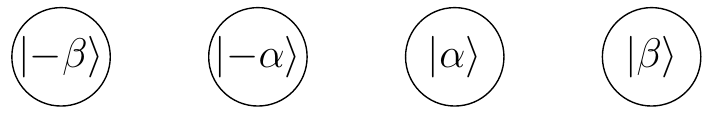}
\caption{Case~3}
\label{subfig:phasespace picture3}
\end{subfigure}
\caption{Illustrative phase-space picture of the linear QTS \ (\ref{eq:QTSeven}) and (\ref{eq:QTSodd}).}
\label{fig:phasespace picture}
\end{figure}

In this subsection,
we define the QTS on the similar lines of idea as the QDS.
We also discuss how the QTS can be
in three different configurations depending on the superposition between the coherent states forming QDSs and the overlap between two such QDSs. This subsection also includes the symmetry transformations of the constituent coherent states and how they change the QTS. 
We define the QTS by extending the concept of QDSs by splitting its two states in two distinguishable states each.
The QTS can be thought of as the superposition of two coherent states separated macroscopically,
each consisting of two coherent states with opposite phases and distinguishable.
The even QTS is
\begin{equation}
\label{eq:QTSeven}
	\ket{\Upsilon}
		\propto\ket{\alpha} +\ket{-\alpha}+\ket{\beta}+\ket{-\beta},
\end{equation}
and the odd QTS is
\begin{equation}
\label{eq:QTSodd}
	\ket{\Upsilon}_\text{odd}
		\propto\ket{\alpha}-\ket{-\alpha}-\ket{\beta}-\ket{-\beta},
\end{equation} 
up to a normalization factor. We study only even QTSs~(\ref{eq:QTSeven})
as the symmetry
\begin{equation}
	\alpha\leftrightarrow\beta,\,
	\alpha\leftrightarrow-\alpha,\,
	\beta\leftrightarrow-\beta,\,
	\alpha\leftrightarrow-\beta,\,
	\alpha\leftrightarrow\beta
\end{equation}
keeps the state the same,
making analysis straightforward,
and we do not lose generality in how to study QTSs.
We consider three different types of QTS~(\ref{eq:QTSeven})
enumerated as Cases~$\Upsilon$1 to~$\Upsilon$3 in~\S\ref{sec:intro}.
For case~$\Upsilon$1,
we have a superposition of two macroscopically separated kitten doublets (Fig.~\ref{subfig:phasespace picture1}).
The second case~$\Upsilon$2,
is a superposition of two macroscopically distinguishable doublets but with the two doublets somewhat overlapping (Fig.~\ref{subfig:phasespace picture2}).
The third case~$\Upsilon$3 is four equally spaced coherent states
on the real line in phase space (Fig.~\ref{subfig:phasespace picture3}).
\subsection{Wigner function and marginal distributions}
\label{subsec:Wigner}
In this subsection,
we describe the Wigner function and marginal distributions for studying the phase-space properties of the QTSs.
In terms of Gaussian functions~(\ref{eq:gaussian}),
the QTS~(\ref{eq:QTSeven}) Wigner function~$W_\text{QTS}(q, p)$
satisfies
\begin{widetext}
\begin{align}
	W_\text{QTS}(q,p;\alpha,\beta)
	=&\frac{1}{\sqrt2N_{\alpha,\beta}} \Bigg[G\left(p;0,1\right)\left[G_+\left(q;\alpha,\frac12\right)
		+G_+\left(q;\beta,\frac12\right)\right]
			\nonumber\\&
		+G\left(q;0,\frac12\right) \bigg[\text{e}^{-2\alpha^2}G_+\left(p;2\text{i}\alpha,1\right)
		+\text{e}^{-2\beta^2}G_+\left(p;2\text{i}\beta,1\right)\bigg]
				\nonumber\\	
	&+\text{e}^{-{(\alpha-\beta)}^2} 
		\bigg[G_+\left(q;\frac{\alpha+\beta}{2},\frac12\right)
			G_+\left(p;\text{i}(\alpha-\beta),1\right)\bigg]
			\nonumber\\&
	+\text{e}^{-{(\alpha+\beta)}^2}
		\bigg[G_+\left(q; \frac{\alpha-\beta}{2},\frac12\right)
			G_+\left(p;\text{i}(\alpha+\beta),1\right)\bigg]\Bigg].
\label{eq:WignerQTS}
\end{align}  
for normalization
\begin{equation}
	N_{\alpha,\beta}
		=2(1+\text{e}^{-2\alpha^2})+2(1+\text{e}^{-2\beta^2})	
			+4\Big(\text{e}^{\frac{-({\alpha-\beta})^2}{2}}+\text{e}^{\frac{-({\alpha+\beta})^2}{2}}\Big)
\label{eq:Normalization_QTS}.
\end{equation}  
\end{widetext}
Now we explain the terms in this complicated expression.

The first term in Eq.~(\ref{eq:WignerQTS}) represents the four Gaussians located at $\pm\alpha$
and~$\pm\beta$ on the phase-space position axis, corresponding to four coherent states.
The second term in~(\ref{eq:WignerQTS}) represents the interference terms between the doublets at~$\ket{\pm\alpha}$ and~$\ket{\pm\beta}$.
The third term in~(\ref{eq:WignerQTS}) corresponds to interference between 
the pair $\ket{\alpha}$ and~$\ket{\beta}$ and the pair $\ket{-\alpha}$ and~$\ket{-\beta}$.
The last term~(\ref{eq:WignerQTS}) shows interference
between the pair $\ket{\alpha}$ and~$\ket{-\beta}$
and the pair~$\ket{-\alpha}$ and~$\ket{\beta}$.

Expression~(\ref{eq:WignerQTS}) readily shows that the interference between each pair of real-amplitude coherent states~$\ket{\pm\alpha}$ and $\ket{\pm\beta}$, which is symmetric about zero on the phase-space position axis,
is located at the centre of the position axis.
However,
interference between the pairs~$\{\ket{\alpha},\ket{\beta}\}$
and $\{\ket{-\alpha},\ket{-\beta}\}$ is centred at $\pm\frac{\alpha+\beta}{2}$,
and,
similarly,
the interference between $\{\ket{\alpha},\ket{-\beta}\}$ and $\{\ket{-\alpha},\ket{\beta}\}$
is centred at $\pm\frac{\alpha-\beta}{2}$, respectively.
Functions of the type $G(p;\text{i}\alpha,\sigma)$
are Gaussian-modulated sinusoidal oscillations along the momentum coordinate of phase space
and represent interference between the coherent states in the QTS.
QTS marginal distributions obtained from the Wigner function following Eq.~(\ref{eq:Wmarginal distributions}) are
\begin{equation}
\label{eq:marginalQTS}
	\operatorname{pr}_\text{QTS}(q)
		=\frac{1}{N_{\alpha,\beta}} \left[G_+\left(q;\alpha,\frac{1}{\sqrt{2}}\right)
			+G_+\left(q;\beta,\frac{1}{\sqrt{2}}\right)\right]^2
\end{equation}
for position, and
\begin{equation}
\label{eq:marginalQTStilde}
	\widetilde{\operatorname{pr}}_\text{QTS}(p)
		=\frac{2\sqrt{2}}{N_{\alpha,\beta}} \Big[G(p;0,1)
			\left(\cos{p\alpha}+\cos{p\beta}\right)^2\Big],
\end{equation}
for momentum.
Equations~(\ref{eq:marginalQTS}) and~(\ref{eq:marginalQTStilde})
can be compared to Eqs.~(\ref{eq:prqcat}) and~(\ref{eq:prpcat}) for the QDS;
evidently, the marginal distribution for QTS along the position axis has four Gaussian peaks whereas the QDS has two Gaussian peaks and the marginal distribution along the momentum axis for QTS has an interference pattern as a Gaussian modulated by
\begin{equation}
\label{eq:cospacospb}
	\left(\cos{p\alpha}+\cos{p\beta}\right)^2,
\end{equation}
which looks like a Cartesian lattice of mountains over~$\mathbb{R}^2$
with coordinates~$\alpha$ and~$\beta$.
For the QDS marginal distribution,
the appearance of the function~(\ref{eq:cospacospb})
as a beat between two tones explains observed interference.
\subsection{Photon-number distribution}
\label{subsec:photonsub}
The photon-number distribution for the QTS is applied for the three cases
$\Upsilon$1 to~$\Upsilon$3
explained in~\S\ref{subsec:definesub}. As an extension of QDSs, photon-number distributions for QTSs are Poissonian distributions corresponding to two QDSs, each of them with an oscillatory nature. Peaks of the oscillation in photon-number distributions are centred at even photon numbers for even QTSs
and at odd photon numbers for odd QTSs as evident in
\begin{equation}
\label{eq:pdQTS}
	\wp^{\pm}_\text{QTS}(n;\alpha,\beta)
		=\left[1\pm(-1)^n\right]
			\tilde{\wp}_\text{QTS}(n;\alpha,\beta)
\end{equation}
for
\begin{align}
\label{eq:pdQTStilde}
	\tilde{\wp}_\text{QTS}(n;\alpha,\beta)
		:=&\frac{4}{N_{\alpha,\beta}}
			\Big[\tilde{\wp}_\text{cs}(n;\alpha,\beta)
							\nonumber\\
				&+\text{e}^{-\frac{(\alpha+\beta)^2}{2}
					+\alpha^2\beta^2}\wp_\text{cs}(n;\alpha\beta)\Big]
\end{align}
and
\begin{equation}
\label{eq:pdQTSsum}
	\tilde{\wp}_\text{cs}(n;\alpha,\beta)
		:=\frac{\wp_\text{cs}(n;\alpha)+\wp_\text{cs}(n;\beta)}{2}
\end{equation}
with~$\wp_\text{cs}(n;\alpha)$
defined in Eq.~(\ref{eq:pndcs}).
As we consider only the case of even QTS,
here we only deal with~$\wp^+_\text{QTS}$ and henceforth drop the superscript~$^+$.
To analyze inter-Poissonian interference we use the expression
\begin{align}
\label{eq:interPoissonianinterference}
	\wp^\text{IP}_\text{QTS}(n;\alpha,\beta)
		=&\frac{4}{N_{\alpha,\beta}}\left[1\pm(-1)^n\right]
					\nonumber\\
			&\times\text{e}^{-\frac{(\alpha+\beta)^2}{2}
				+\alpha^2\beta^2}\wp_\text{cs}(n;\alpha\beta)
\end{align}
with~$^+$ suppressed and with the superscript~$^\text{IP}$ referring to `inter-Poissonian'.

The first two terms in (\ref{eq:pdQTS}) indicate that the two Poissonian curves for each QDS consist of two coherent states with the same amplitude but opposite phases, e.g., $\ket{\pm\alpha}$ and~$\ket{\pm\beta}$ are peaked at $\left|\alpha\right|^2$ and~$\left|\beta\right|^2$, respectively.
The third term on the right-hand side of Eq.~(\ref{eq:pdQTS})
represents a small interference pattern
between these two Poissonian distributions.

The inter-Poissonian interference term leading is intriguing and needs to be established as interference between distinct Poissonian sectors or not. 
To understand the inter-Poissonian interference term, we analyze the photon-number distribution involving only the inter-Poissonian interference term in Eq.~(\ref{eq:pdQTS}).
To elucidate the interference effect, we take the derivative of the envelope photon-number distribution with respect to photon number~$n$
by treating~$n$ as a continuous quantity hence using the relation
$n!=\Gamma(n+1)$.

The photon-number distribution~(\ref{eq:pdQTS}) 
has an envelope function~(\ref{eq:pdQTStilde}),
which is the sum of Poissonian distributions~(\ref{eq:pdQTSsum})
plus the second-term on the right-hand side of Eq.~(\ref{eq:pdQTStilde})
corresponding to inter-Poissonian interference.
Now we take the derivatives of the envelope function including inter-Poissonian interference~(\ref{eq:pdQTStilde})
and excluding inter-Poissonian interference~(\ref{eq:pdQTSsum}).
The derivative of the envelope function reveals the envelope function's extrema as zeroes,
which makes it easy to identify the extrema and especially whether the extrema change due to inter-Poissonian interference.

We take derivatives of this envelope function (\ref{eq:pdQTStilde}) which includes the inter-Poissonian interference 
\begin{align}
\label{eq:pd'QTS_1}
	&\tilde{\wp}'_\text{QTS}(n;\alpha,\beta)\nonumber\\
		=&\frac{N^{-1}_{\alpha,\beta}}{\Gamma(n+1)} \Bigg[2 \text {e}^{-(\alpha^2+\beta^2)}\Big( \alpha^n\text{e}^\frac{\beta^2}{2}+\beta^n\text{e}^\frac{\alpha^2}{2}\Big)
				\nonumber\\
		&\times\Big[2 \Big(\alpha^n\text{e}^\frac{\beta^2}{2}\log\alpha+\beta^n\text{e}^\frac{\alpha^2}{2}\log\beta\Big) \nonumber\\
		&-\Big(\alpha^n\text{e}^\frac{\beta^2}{2}+\beta^n\text{e}^\frac{\alpha^2}{2}\Big) \Psi^{(0)}(n+1)\Big]\Bigg],
\end{align}
and the sum~(\ref{eq:pdQTSsum}) which does not include the inter-Poissonian interference

\begin{align}
\label{eq:pd'QTS_2}
                 &\tilde{\wp}'_\text{cs}(n;\alpha,\beta)\nonumber\\
		=&\frac{N^{-1}_{\alpha,\beta}}{\Gamma(n+1)} \Bigg[ \text {e}^{-(\alpha^2+\beta^2)}\nonumber\\
		&\times\Big[4 \Big(\alpha^{2n}\text{e}^{\beta^2}\log\alpha+\beta^{2n}\text{e}^{\alpha^2}\log\beta\Big) \nonumber\\
		&-\Big(\alpha^{2n}\text{e}^{\beta^2}+\beta^{2n}\text{e}^{\alpha^2}\Big) \Psi^{(0)}(n+1)\Big]\Bigg].
\end{align}
The sign~$'$ signifies differentiation with respect to~$n$ and
\begin{equation}
\label{eq:digamma}
	\Psi^{(0)}(n)
		:=\frac{\text{d}}{\text{d}n}\ln(\Gamma(n))
\end{equation}
is the digamma function.
\subsection{Multiple-well approximation}
\label{multiplewell approx}
The idea of approximately realizing a QDS as a double-well ground state is discussed in~\S\ref{subsec:doublewellpotential}.
Similarly,
we propose a QTS implementation based on realizing the ground state
of a quadrupole-Gaussian-well potential under reasonable approximations.

We study a quadruple Gaussian potential well structures
\begin{align}
\label{eq:VxquadrupleGaussian}
	V(x)
		=&V_\text{g}(x-\alpha)
			+V_\text{g}(x+\alpha)
				+V_\text{g}(x-\beta)
						\nonumber\\
			&+V_\text{g}(x+\beta)
				-V_\text{g}(0),
\end{align}
where
\begin{equation}
	V_\text{g}(x)
		=-V_0 \text{exp}\left(-\frac{\gamma x^2}{2\sigma^2}\right)
\end{equation}
is a single-well Gaussian potential centred at $x=0$.
The ground state of the single-particle Hamiltonian with potential~(\ref{eq:VxquadrupleGaussian})
serves as an approximation to the QTS.
Our multiple-well potential~(\ref{eq:VxquadrupleGaussian})
differs from Nandi's approach to creating a double-Gaussian-well potential~\cite{Nan11},
as discussed in~\S\ref{subsec:doublewellpotential}
as Nandi effectively uses a Hermite-Gaussian potential well whereas we employ a sum of Gaussian wells
with varied spacing to treat each QTS case described in~\S\ref {subsec:definesub}.
For numerical simulation,
our Gaussian potential wells are approximated by piecewise-continuous functions.

Our method involves transforming a Schr{\"o}dinger equation into a matrix equation from a differential equation by using the three-point finite difference method.
The matrix eigenvalue problem is solved numerically by using the inverse iteration method to obtain the Wigner function of a ground-state vector.
\section{Analysis and results}
\label{sec:analysis and results}
In this section,
we refer to the pairs of two coherent states $\ket{\alpha}$, $\ket{\beta}$ and $\ket{-\alpha},\ket{-\beta}$ as doublets and analyze three different cases of the QTS~(\ref{eq:QTSeven})
based on the distinguishability between the doublet states and the states of doublets according to the three specific QTS cases
\begin{itemize}
	\item[$\Upsilon$1:]
		$\alpha=4$, $\beta=7$
			(``two doublets''),
	\item[$\Upsilon$2:]
		$\alpha=1,\beta=6$
			(``doublet between two singlets''),
	\item[$\Upsilon$3:]
		$\alpha=2,\beta=6$
			(``comb state'').
\end{itemize}
The first case describes the configuration in which doublets are separated by a macroscopic distance, but the coherent states forming the doublets are close.
The second case describes the configuration in which two doublets are close together, but the coherent states forming the doublets are macroscopically separated. The third case is a depiction of the QTS where all the coherent states are equally spaced with macroscopic distinguishability.
We apply our four tools comprising the Wigner function, 
marginal distributions,
photon-number distribution
and finally the multiple-well approximation using Wigner function. 
\subsection{$\Upsilon1$: Two doublets}
\label{subsec:Upsilon1}
In this subsection,
we analyze the two-doublet QTS ($\Upsilon1$).
Specifically,
we analyze the results for the Wigner function,
the marginal distributions,
the photon-number distribution,
and the four-well ground-state approximation.
\subsubsection{$\Upsilon1$: Wigner function}
\begin{figure}
\centering
\begin{subfigure}{0.40\textwidth}
\includegraphics[width=\linewidth]{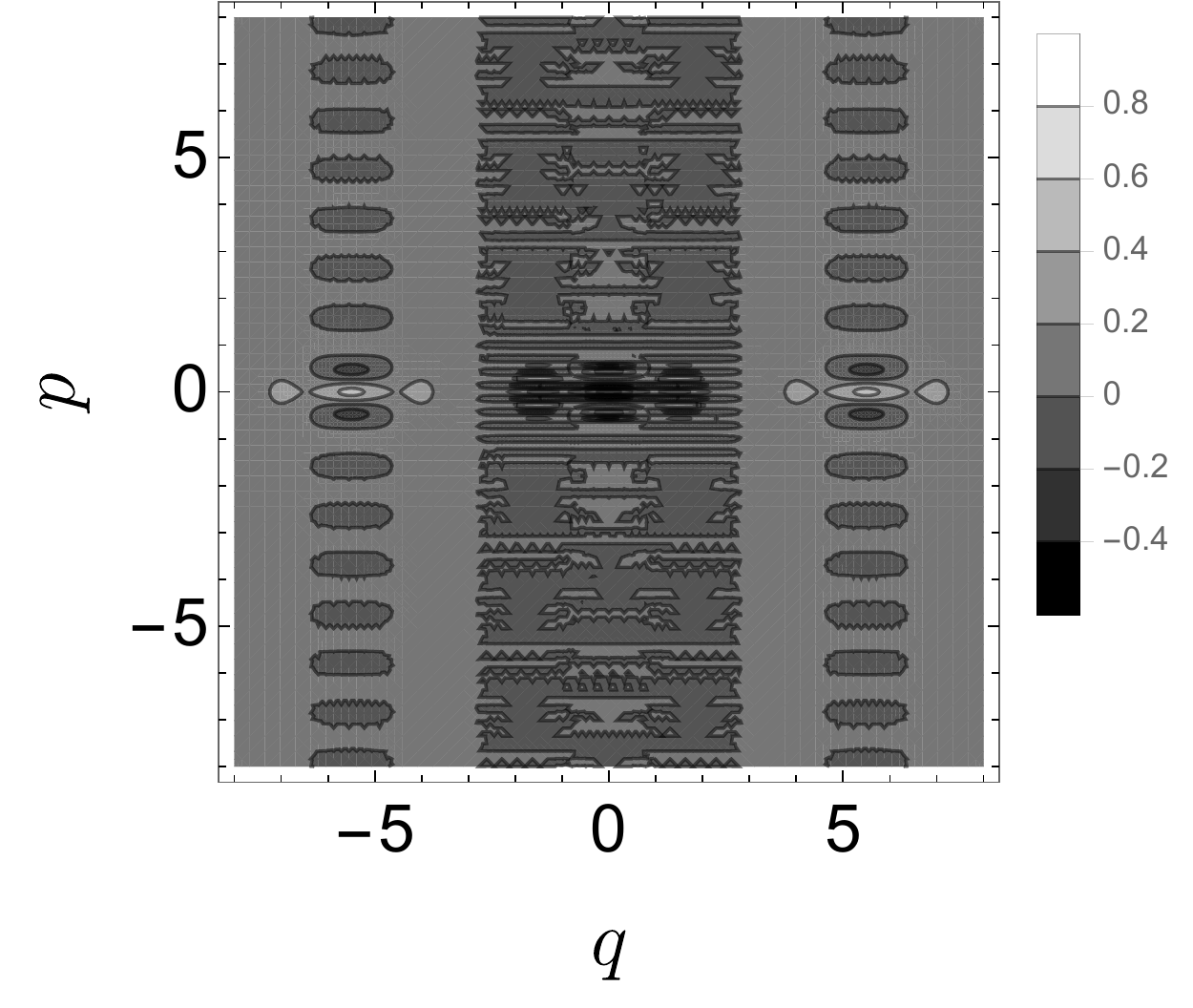} 
\caption{ }
\label{subfig:Wignermarginal1_1}\vspace{0.5 cm}
\end{subfigure}
\begin{subfigure}{0.23\textwidth}
\includegraphics[width=\linewidth]{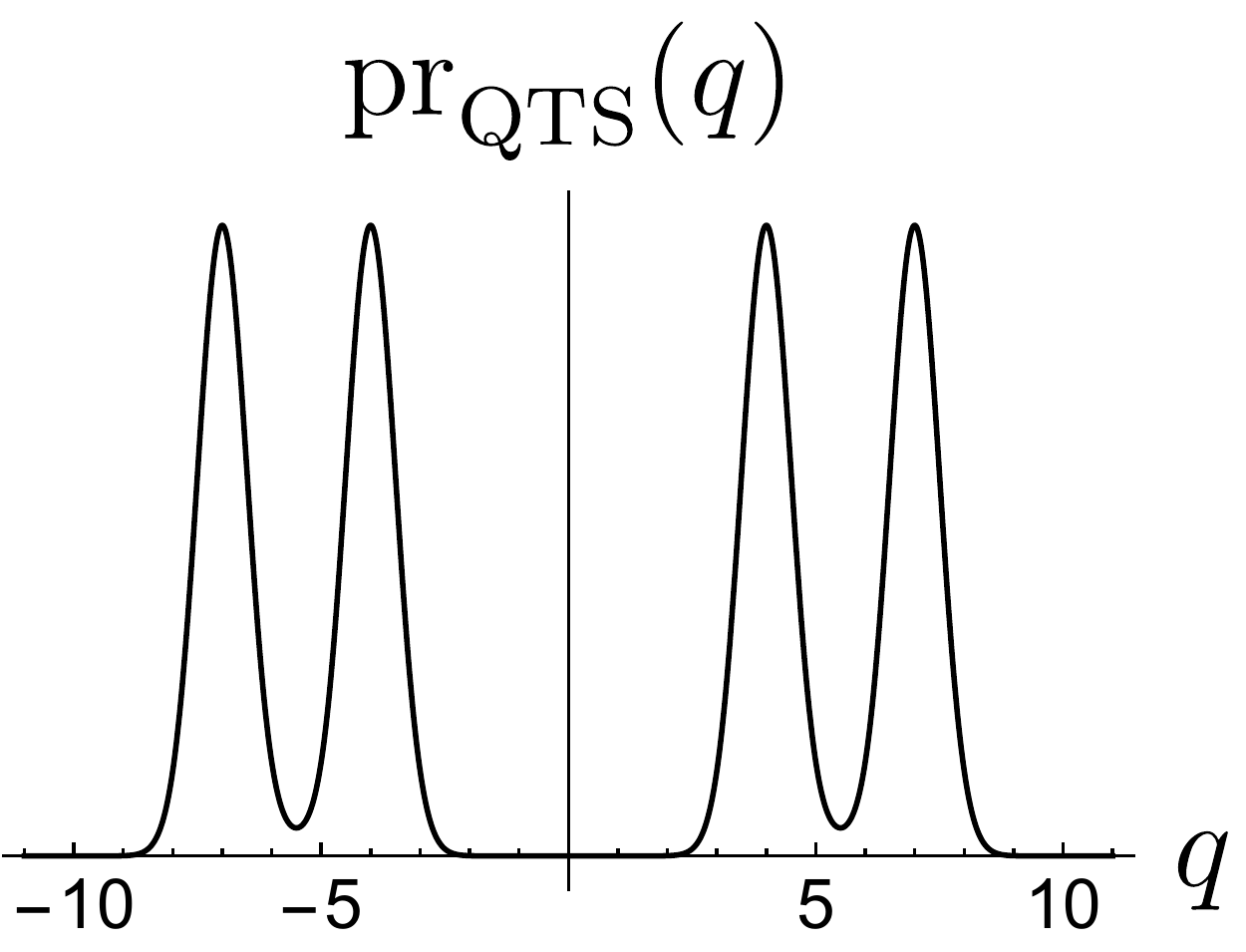}
\caption{ }
\label{subfig:Wignermarginal1_2}
\end{subfigure}
\hspace{0.1 cm}
\begin{subfigure}{0.23\textwidth}
\includegraphics[width=\linewidth]{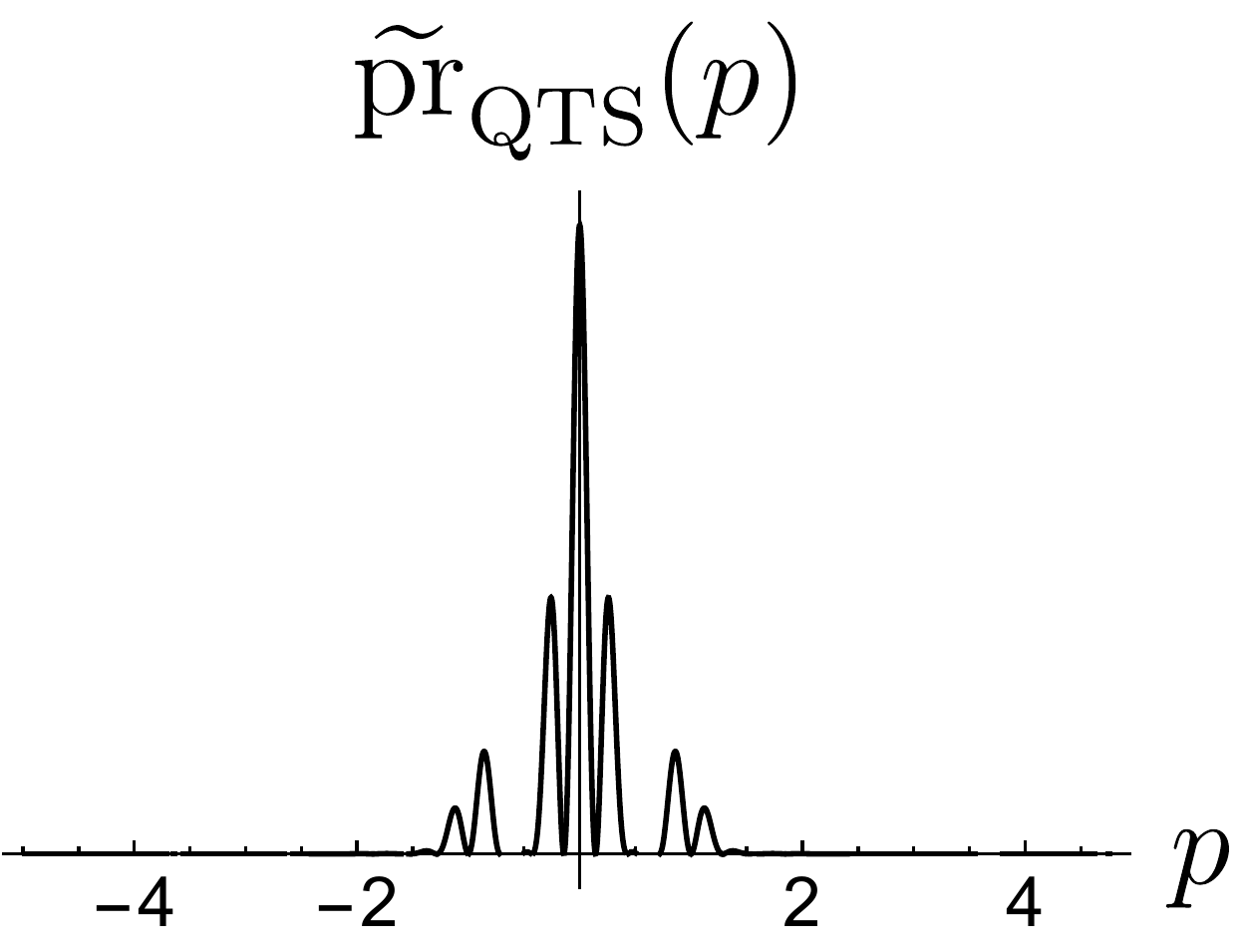}
\caption{ }
\label{subfig:Wignermarginal1_3}
\end{subfigure} 
 \caption{Wigner function and QTS marginal distributions case~$\Upsilon1$ for $\alpha=4,\beta=7$.}
 \label{fig:Wignermarginal1}
\end{figure}

We present the Wigner function for the~$\Upsilon1$ QTS in Fig.~\ref{subfig:Wignermarginal1_1}.
In this figure,
we observe four Gaussian peaks in contrast to the pair of Gaussian peaks for the QDS~(\ref{eq:prqcat})
shown in Fig.~\ref{fig:Wcat}.
Whereas the QDS has a sinusoidal oscillation~(\ref{eq:prpcat})
seen in Fig.~\ref{fig:Wcat}, 
the QTS has an oscillatory beat in momentum~(\ref{eq:marginalQTStilde})
seen in Fig.~\ref{subfig:Wignermarginal1_1}.
For this QTS arrangement, as we can infer from the last section and Eq.~(\ref{eq:WignerQTS}), the four flattened Gaussians located at $\pm4$ and at $\pm7$
correspond to the coherent state~$\ket{\alpha}$ for $\alpha\in\{\pm4,\pm7\}$.

The interference patterns seen between the doublets are coming from the states~$\ket4$, $\ket{7}$
and~$\ket{-7}$, $\ket{-4}$ centred in the middle of each pair, i.e., $5.5$ and $-5.5$, respectively.
The interference at the origin is caused by $\ket{\pm4}$ and $\ket{\pm7}$, on its right due to the pair~$\ket{-4}$ and~$\ket{7}$,
and on its left due to the pair~$\ket{4}$ and~$\ket{-7}$ centred at~$1.5$ and~$-1.5$, respectively, on the position axis. The two interference patterns centred at $1.5$ and~$-1.5$ are not as distinguishable as for QDSs.
\subsubsection{$\Upsilon1$: Marginal distributions}
QTS marginal distributions for~$\Upsilon1$ are shown in Figs.~\ref{subfig:Wignermarginal1_2}  and~\ref{subfig:Wignermarginal1_3}
for position and momentum variables and are obtained by integrating the Wigner function  over the conjugate degree of freedom.
In Fig.~\ref{subfig:Wignermarginal1_2},
we see four Gaussian peaks and no interference for the marginal distribution along the position axis.
However, the marginal distribution along the momentum axis in Fig.~\ref{subfig:Wignermarginal1_3} shows an interference pattern with a Gaussian modulated by the square of sum of two sinusoidal functions,
i.e., $(\cos4p+\cos7p)^2$.
 
\subsubsection{$\Upsilon1$: Photon-number distribution}
\begin{figure}
\centering
\begin{subfigure}{0.40\textwidth}
\includegraphics[width=\linewidth]{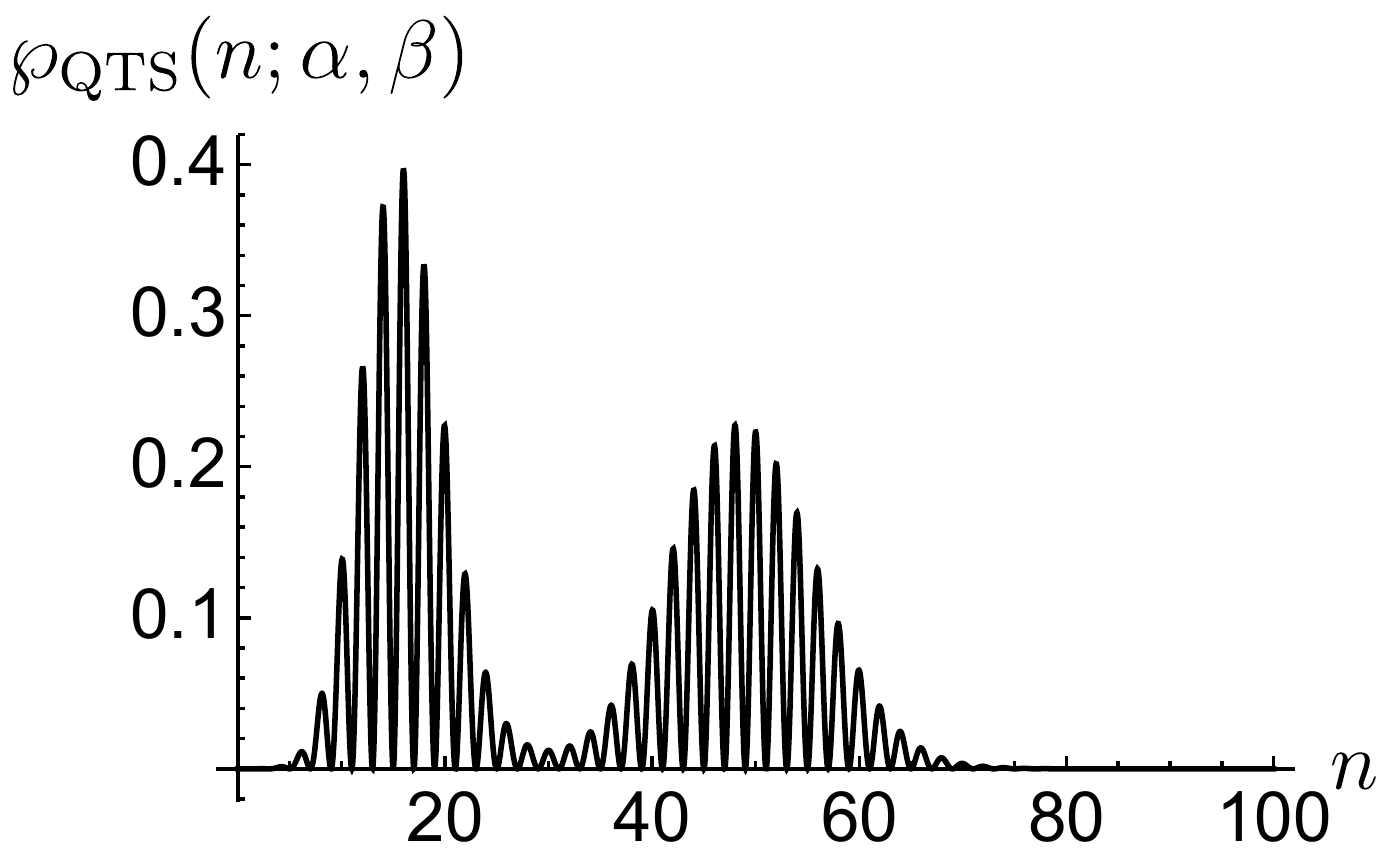} 
\caption{ }
\label{subfig:pndevenqtscase1_1}\vspace{0.5 cm}
\end{subfigure}
\begin{subfigure}{0.23\textwidth}
\includegraphics[width=\linewidth]{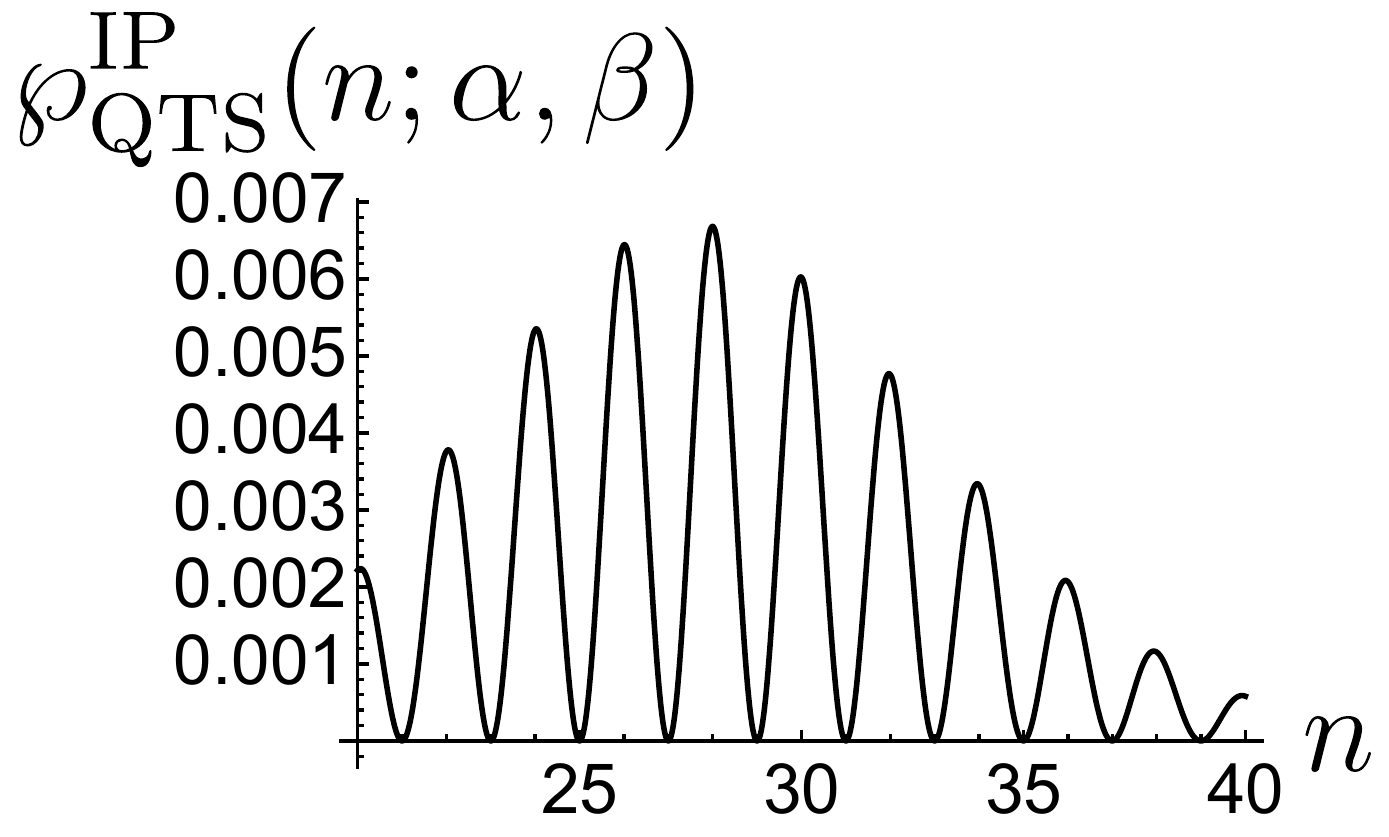}
\caption{ }
\label{subfig:pndevenqtscase1_2}
\end{subfigure}
\hspace{-0.1 cm}
\begin{subfigure}{0.23\textwidth}
\includegraphics[width=\linewidth]{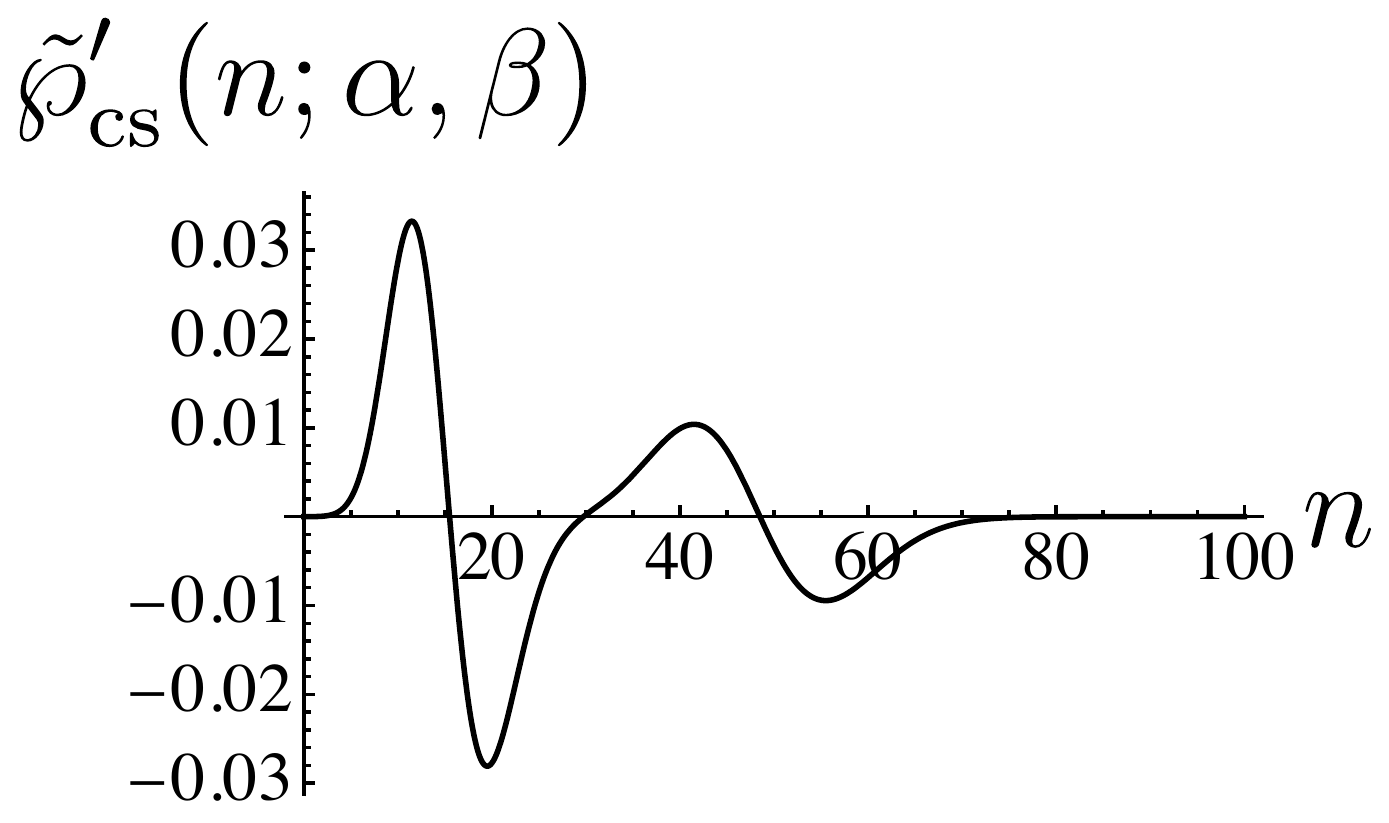}
\caption{ }
\label{subfig:pndevenqtscase1_3}
\end{subfigure}
\caption{%
	$\Upsilon1$:
	Plots of
	\subref{subfig:pndevenqtscase1_1}
	$\wp_\text{QTS}(n;\alpha,\beta)$,
	\subref{subfig:pndevenqtscase1_2}
	$\wp^\text{IP}_\text{QTS}(n;\alpha,\beta)$
	and
	\subref{subfig:pndevenqtscase1_3}
	$\tilde{\wp}'_\text{cs}(n;\alpha,\beta)$
	for $\alpha=4$ and~$\beta=7$.%
}
\label{fig:pndevenqtscase1}
\end{figure}

For the same example of~$\Upsilon1$,
we plot photon-number distribution $\wp_\text{QTS}(n;\alpha,\beta)$~(\ref{eq:pdQTS})
in Fig.~\ref{subfig:pndevenqtscase1_1},
which shows two modulated Poissonian distributions corresponding to the two QDSs $\ket{\pm4}$ and $\ket{\pm7}$ with peaks located at $16=(\pm4)^2$ and~$49=(\pm7)^2$
corresponding to the amplitude peaks of the QTS. The zoomed view of interference effect between these two Poissonian distributions~(\ref{eq:interPoissonianinterference}) is evident in Fig.~\ref{subfig:pndevenqtscase1_2}. Figure \ref{subfig:pndevenqtscase1_3} shows the derivative of envelope photon-number distribution without inter-Poissonian interference term given in Eq.~(\ref{eq:pd'QTS_2}).
From Fig.~\ref{subfig:pndevenqtscase1_3}, it is clear that the interference is too small to affect the infimum of the envelope curve in (\ref{eq:pdQTStilde}). 
\subsubsection{$\Upsilon1$: Quadruple-Gaussian-well ground state}
\begin{figure}
\centering
\begin{subfigure}{0.40\textwidth}
\includegraphics[width=\linewidth]{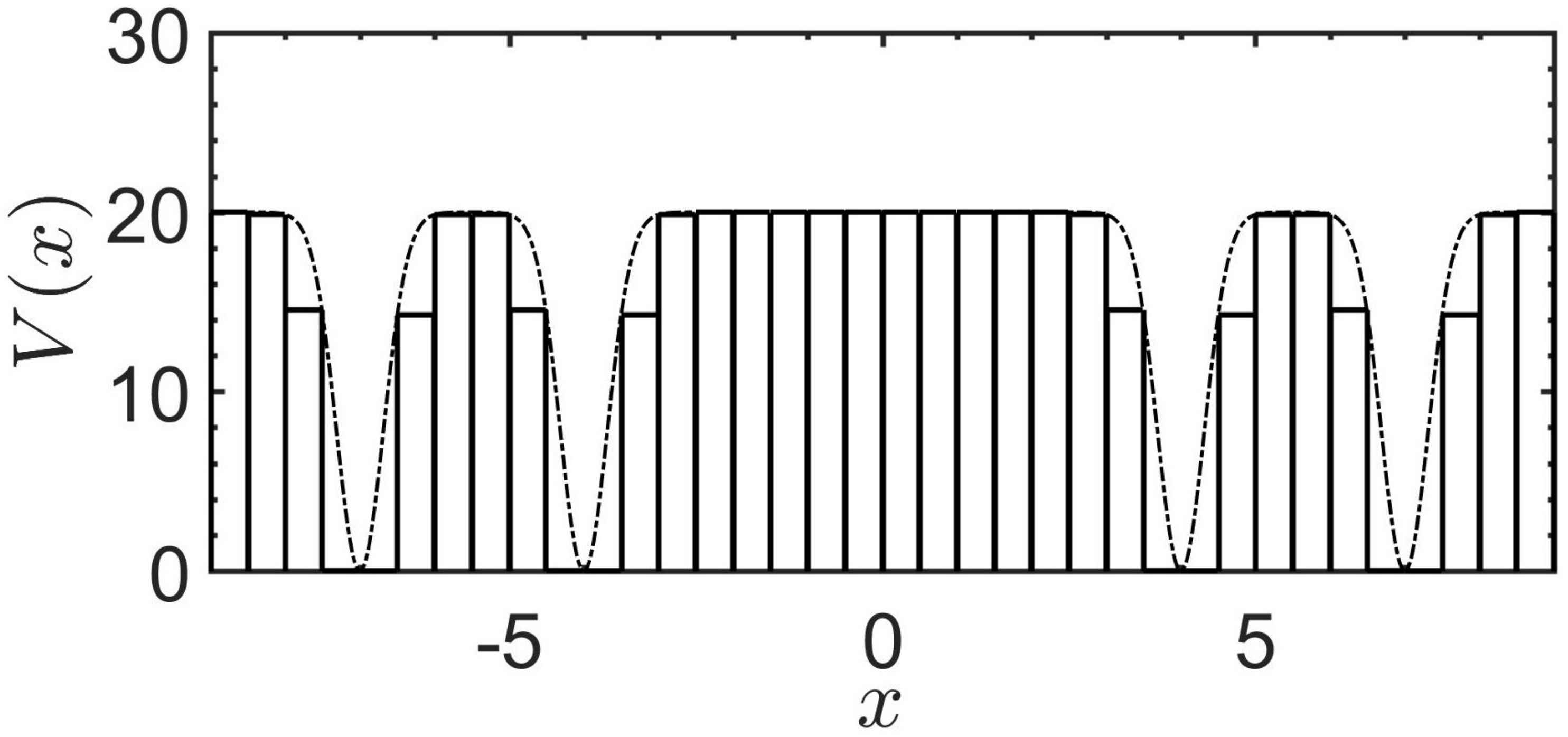}
\caption{ }
\label{subfig:gaussianwellscase1_1}
\end{subfigure}
\vspace*{-0.1cm}
	\hspace*{3cm}
  \flushleft
\begin{subfigure}{0.44\textwidth}
\hspace*{0.5cm}
\includegraphics[width=\linewidth]{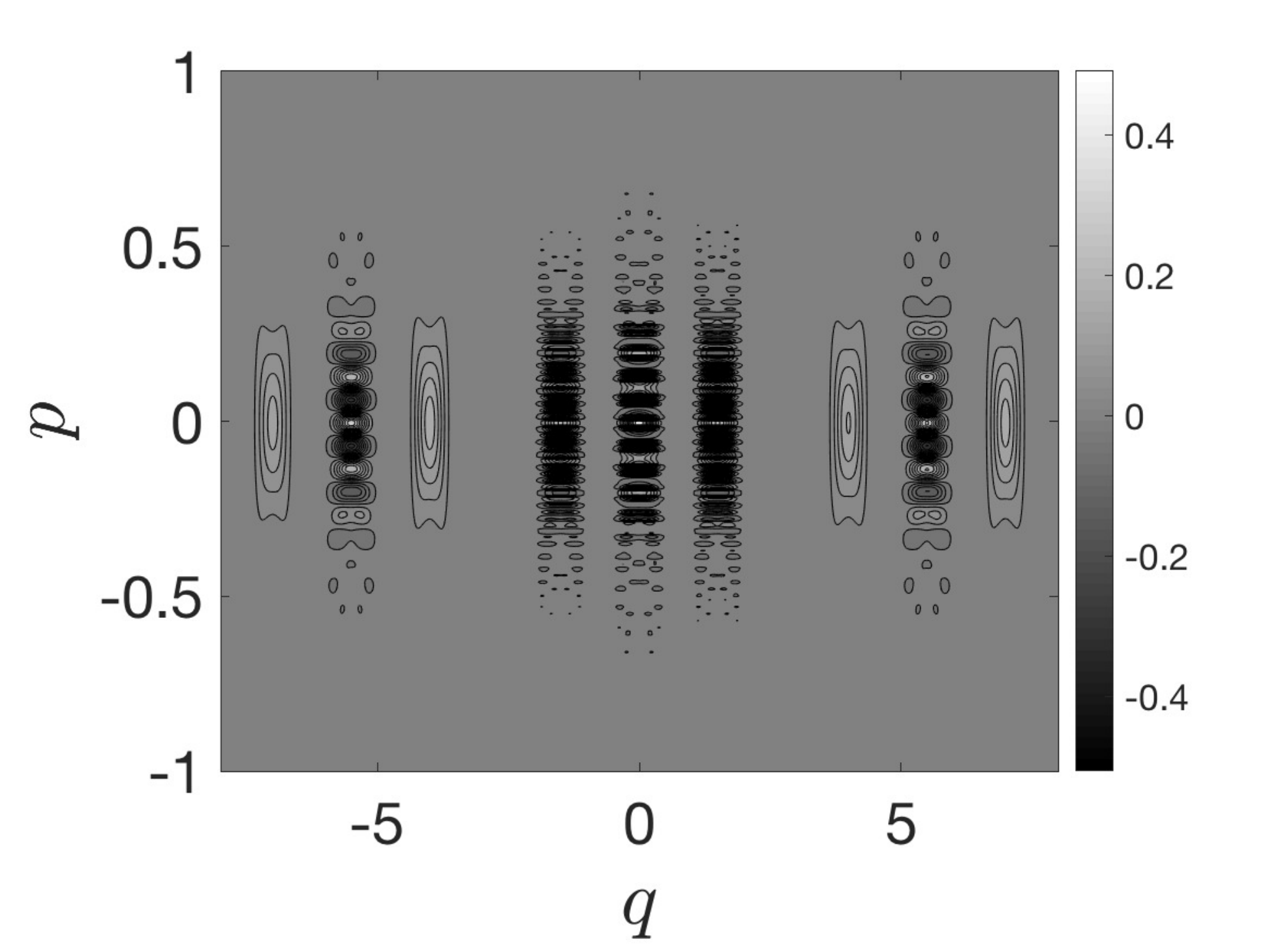}
\caption{ }
\label{subfig:gaussianwellscase1_2}
\end{subfigure}
\caption{$\Upsilon1$:~\subref{subfig:gaussianwellscase1_1}
	quadrupole-Gaussian-well potential approximated 
	as a piecewise continuous function and~\subref{subfig:gaussianwellscase1_2} ground-state Wigner function for the potential $V(x)$ for $\alpha=4,\beta=7$ ($\Upsilon1$).}
\label{fig:gaussianwellscase1}
\end{figure}

We calculate the Wigner function for the approximated ground state of the quadruple-Gaussian-well potential in this configuration shown in Fig.~\ref{subfig:gaussianwellscase1_1}.
The Wigner function for this ground state is plotted in Fig.~\ref{subfig:gaussianwellscase1_2},
and the pattern of the Wigner function closely matches the analytic Wigner function for this QTS
configuration in Fig.~\ref{subfig:Wignermarginal1_1}.
This close match holds in the sense that the Gaussian peaks and interference effects are at the same places in both plots for the $p=0$ axis.
Furthermore, the interference effects proceed ad infinitum.
Minor differences between the plots are expected because the Gaussian wells only approximate parabolic potentials,
which would be needed to see very close approximations to coherent states in the superposition.
\subsection{$\Upsilon2$: Doublet between two singlets}
\label{subsec:Upsilon2}
In this subsection,
we analyze the doublet-between-two-singlets QTS ($\Upsilon2$).
Specifically,
we analyze the results for the Wigner function,
the marginal distributions,
the photon-number distribution,
and the four-well ground-state approximation.
\subsubsection{$\Upsilon2$: Wigner function}
\begin{figure}
\centering
\begin{subfigure}{0.40\textwidth}
\includegraphics[width=\linewidth]{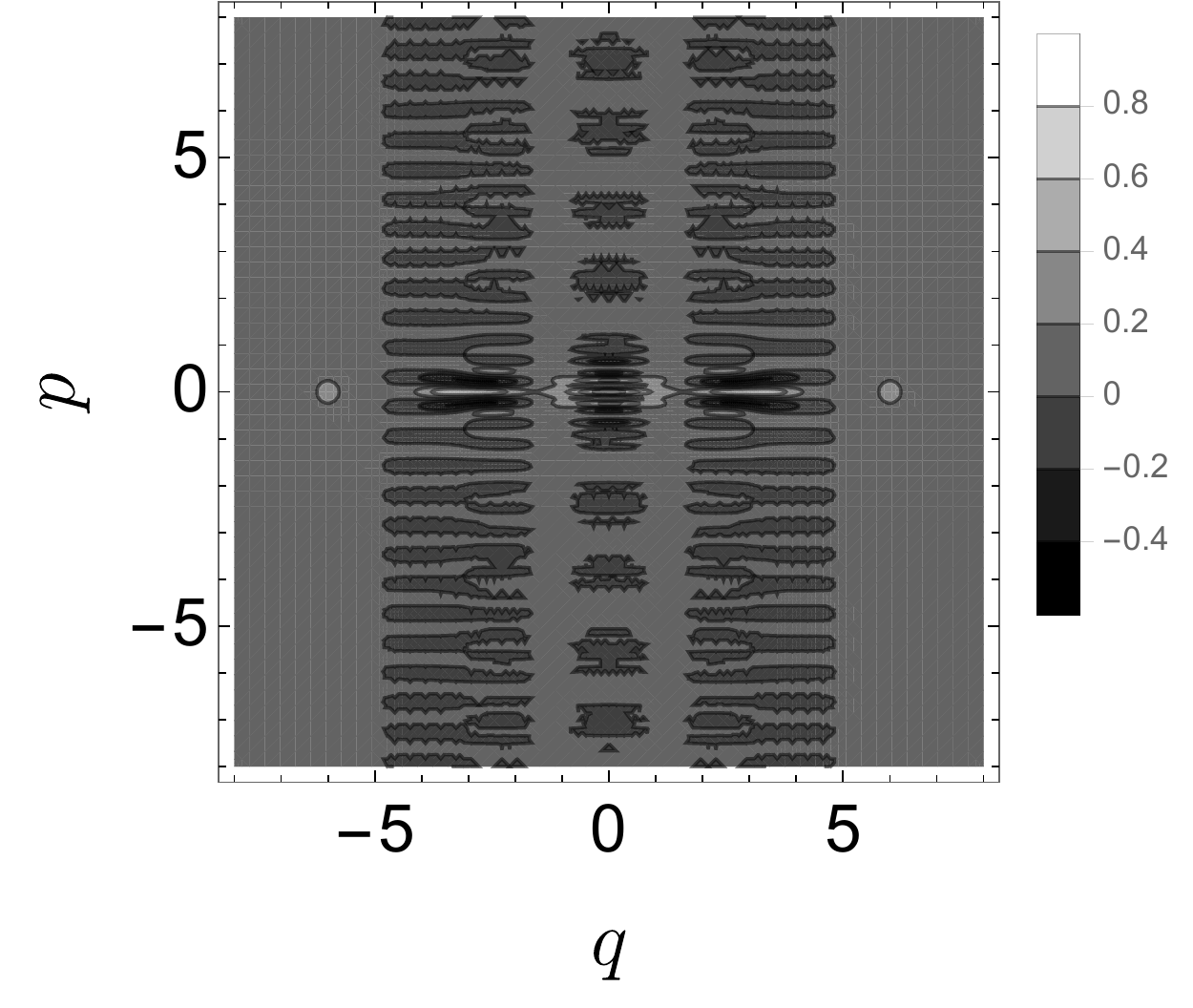} 
\caption{ }
\label{subfig:Wignermarginal2_1}\vspace{0.5 cm}
\end{subfigure}
\begin{subfigure}{0.23\textwidth}
\includegraphics[width=\linewidth]{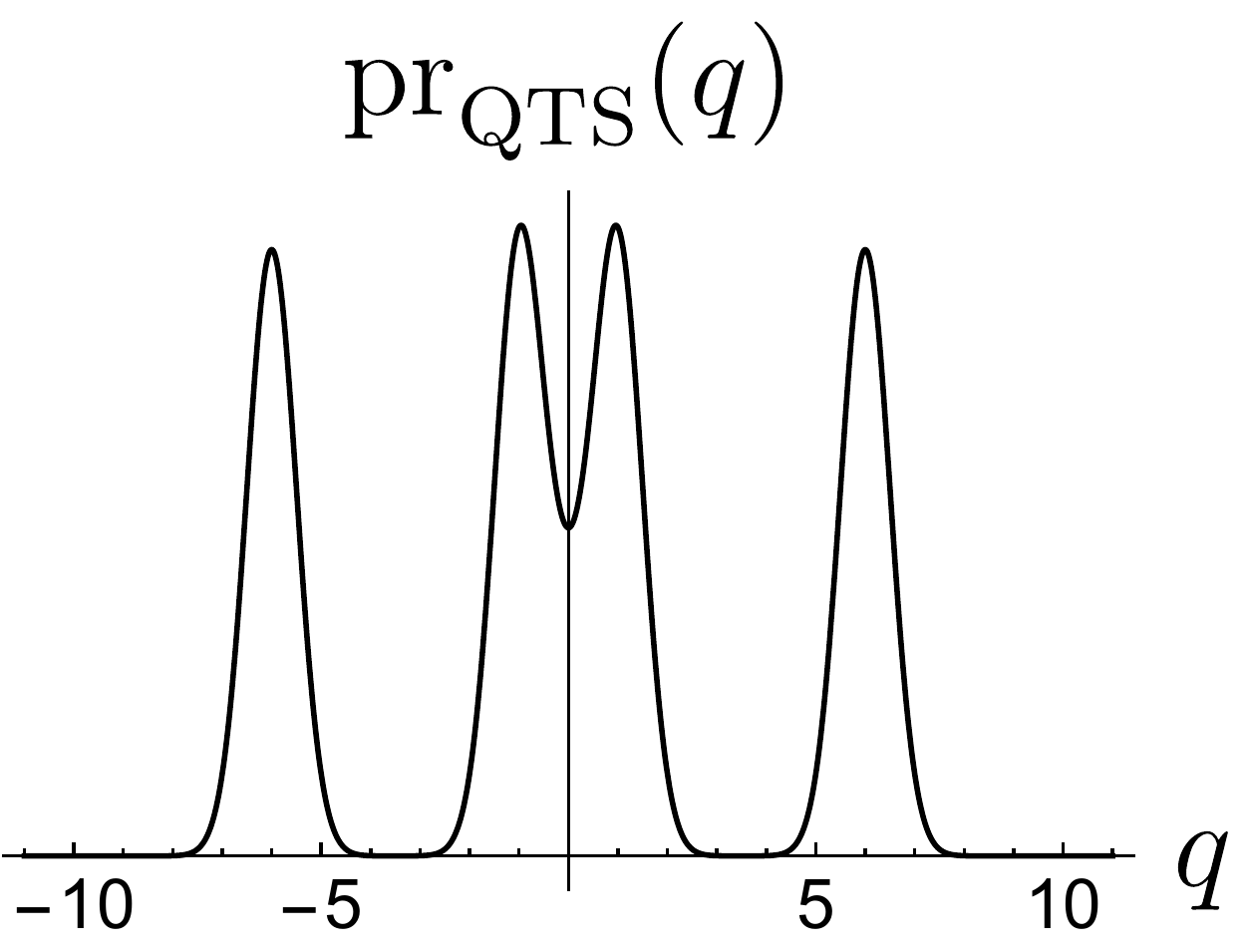}
\caption{ }
\label{subfig:Wignermarginal2_2}
\end{subfigure}
\hspace{0.1 cm}
\begin{subfigure}{0.23\textwidth}
\includegraphics[width=\linewidth]{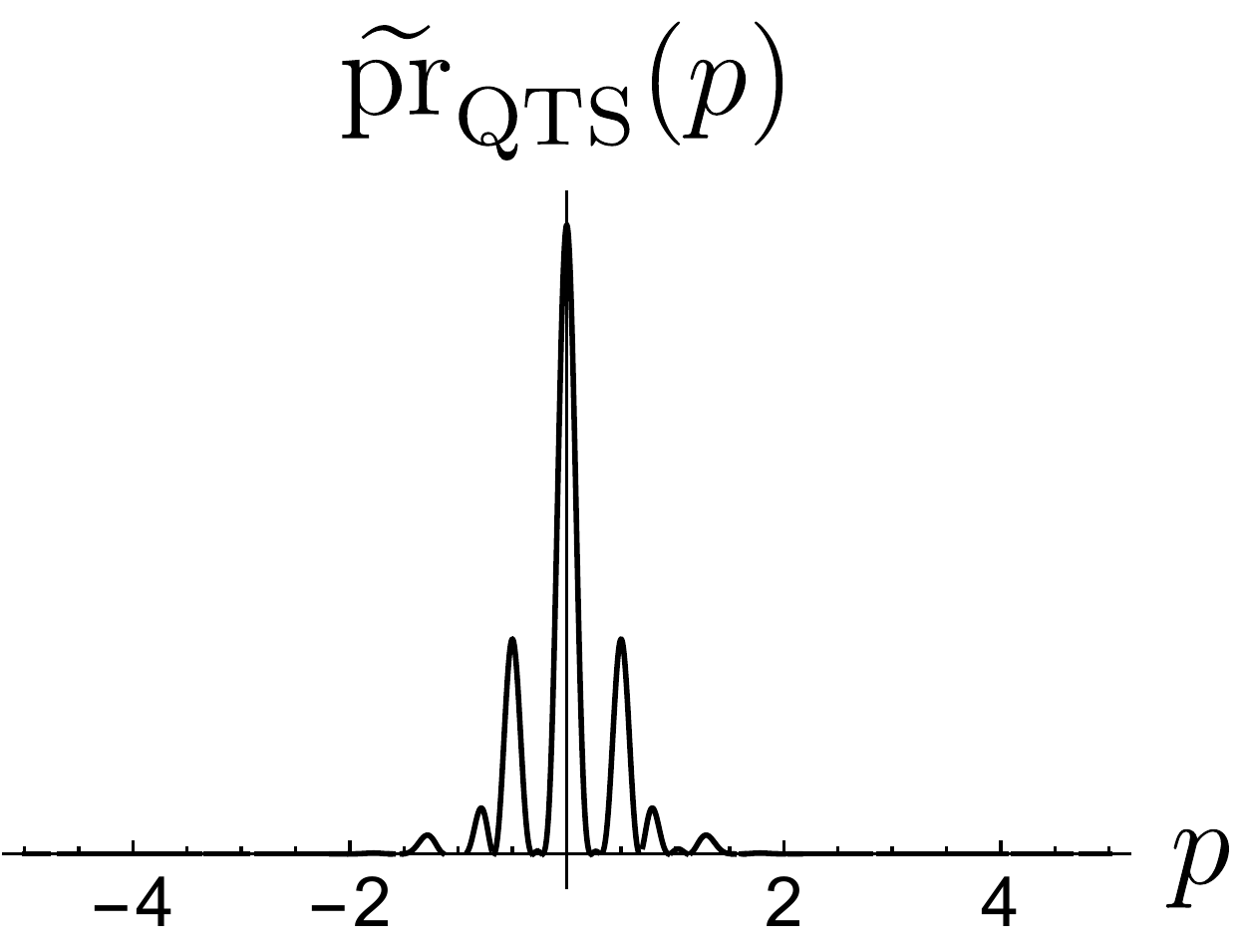}
\caption{ }
\label{subfig:Wignermarginal2_3}
\end{subfigure} 
\caption{Wigner function and QTS marginal distributions case~$\Upsilon2$ for  $\alpha=1,\beta=6$.}
 \label{fig:Wignermarginal2}
\end{figure}

We present the Wigner function for the~$\Upsilon2$ QTS in Fig.~\ref{subfig:Wignermarginal2_1}.
The locations of the Gaussians corresponding to states~$\ket{\pm1}$ and~$\ket{\pm6}$ on the position axis are $\pm1$ and~$\pm6$. The interference pattern at the origin is due to the coherent states~$\ket{\pm1}$ and $\ket{\pm6}$ but the states~$\ket{\pm1}$ are too close to form a QDS and the Gaussians corresponding to these states run into each other as in Fig.~\ref{subfig:Wcat2}. This proximity between states represented in phase space causes the Gaussian peaks at $\pm1$ not showing explicitly and leading to the interference pattern between $\ket{-6}, \ket1$ centred at $-2.5$ overlap with $\ket{-6}, \ket{-1}$ at $-3.5$ and $\ket6, \ket{-1}$ at $2.5$ overlap with $\ket6, \ket1$ at $3.5$. These overlaps cause the interference patterns to spread along the $q$ axis.
\subsubsection{$\Upsilon2$: Marginal distributions}
The QTS marginal distributions for configuration~$\Upsilon2$ are shown in Figs.~\ref{subfig:Wignermarginal2_2} and~\ref{subfig:Wignermarginal2_3}.
We see the Gaussian peaks at $\pm1$ and $\pm6$ along the position axis and an interference Gaussian modulated by $(\cos p+\cos6p)^2$ along the momentum quadrature. 
\subsubsection{$\Upsilon2$: Photon-number distribution}
\begin{figure}
\centering
\begin{subfigure}{0.40\textwidth}
\includegraphics[width=\linewidth]{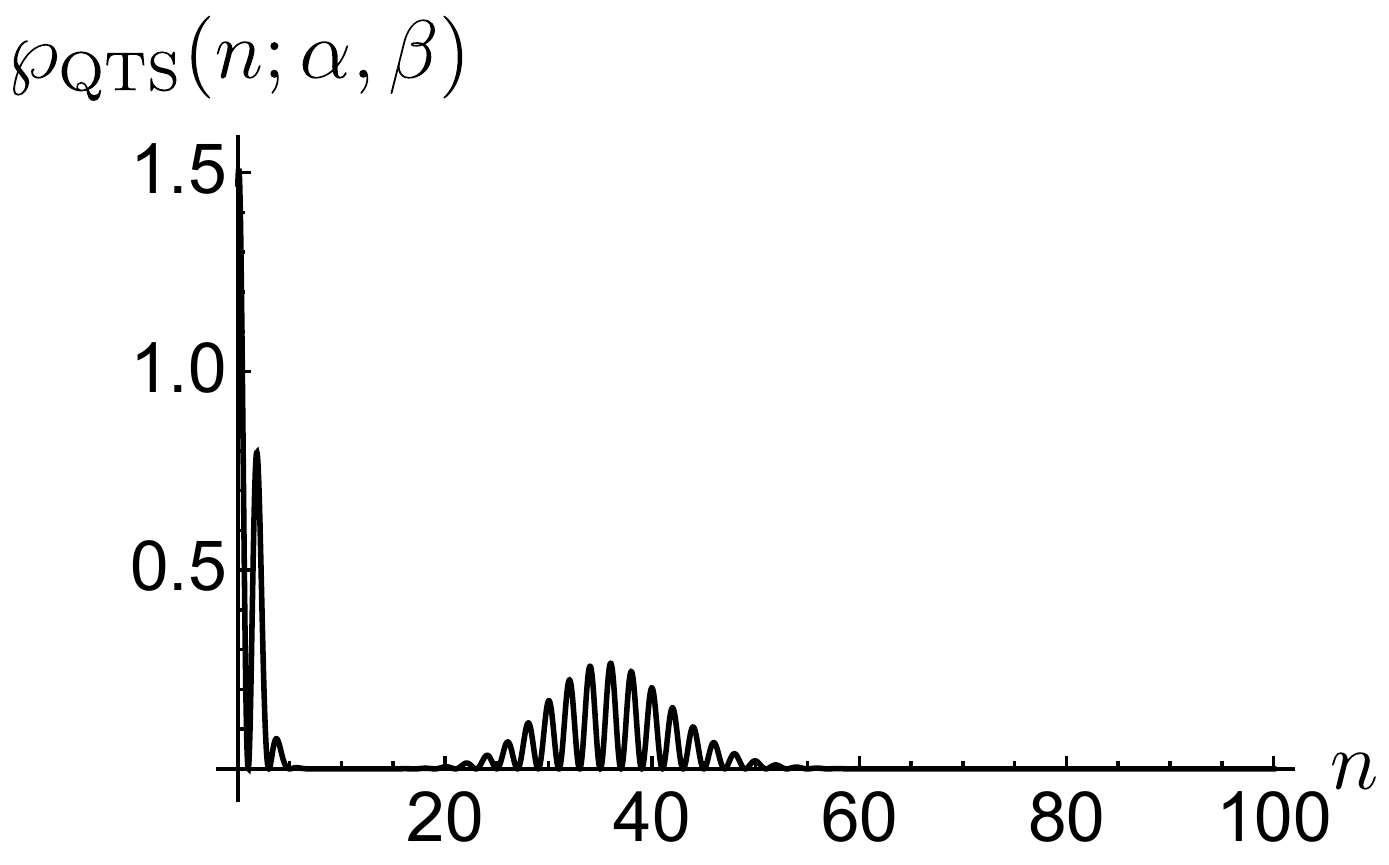} 
\caption{ }
\label{subfig:pndevenqtscase2_1}
\vspace{0.5 cm}
\end{subfigure}
\begin{subfigure}{0.23\textwidth}
\includegraphics[width=\linewidth]{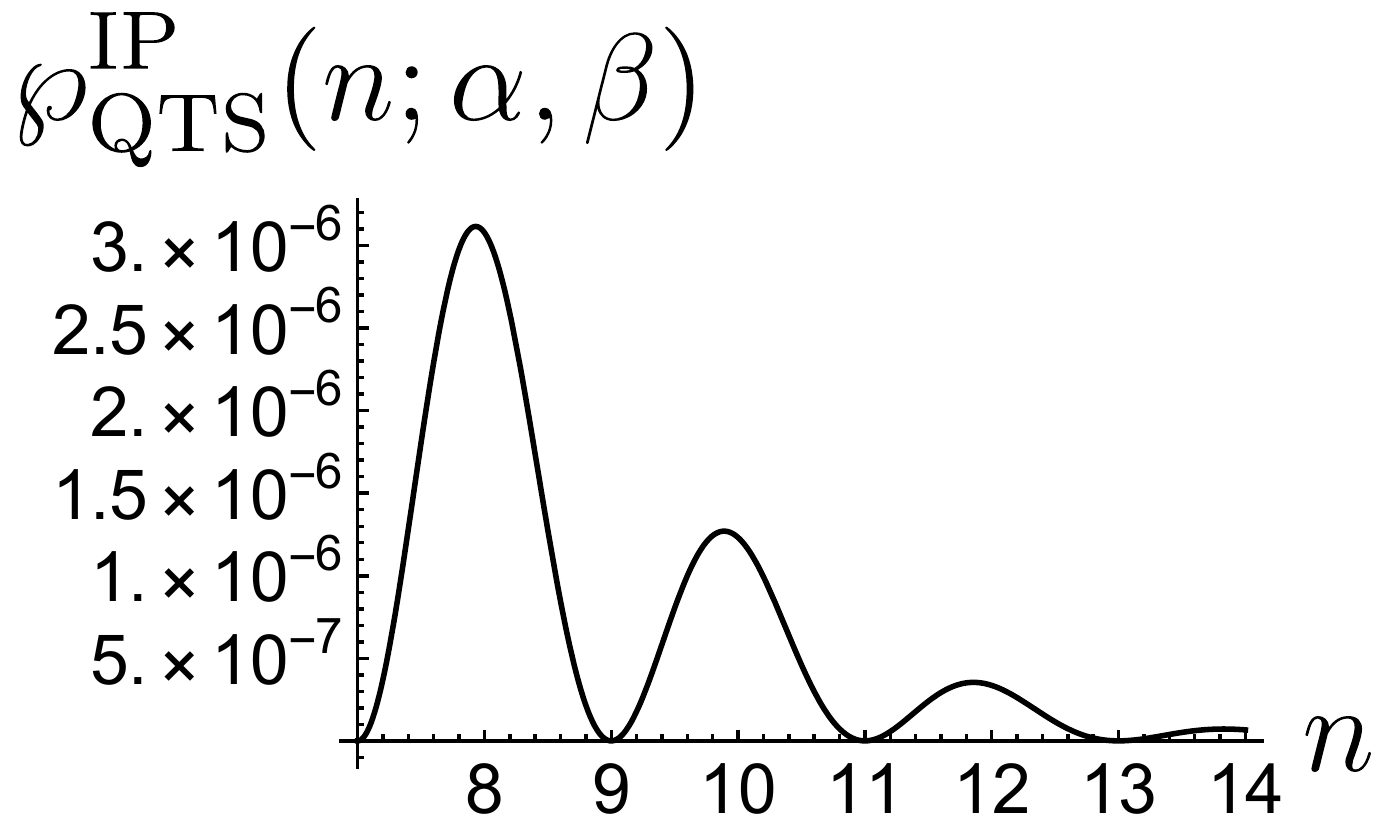}
\caption{ }
\label{subfig:pndevenqtscase2_2}
\end{subfigure}
\hspace{-0.1 cm}
\begin{subfigure}{0.23\textwidth}
\includegraphics[width=\linewidth]{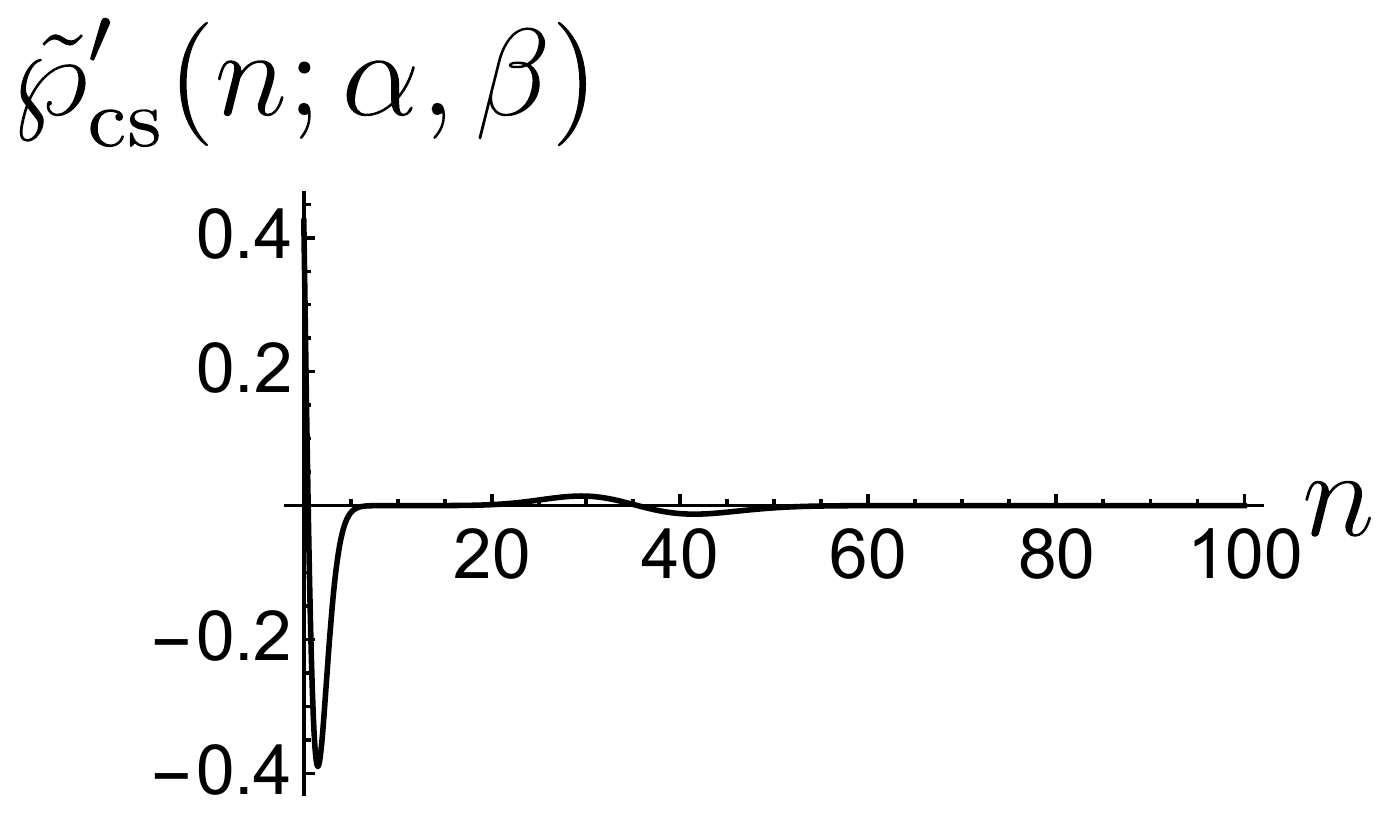}
\caption{ }
\label{subfig:pndevenqtscase2_3}
\end{subfigure}
\caption{%
	$\Upsilon2$:
	Plots of
	\subref{subfig:pndevenqtscase2_1}
	$\wp_\text{QTS}(n;\alpha,\beta)$,
	\subref{subfig:pndevenqtscase2_2}
	$\wp^\text{IP}_\text{QTS}(n;\alpha,\beta)$
	and
	\subref{subfig:pndevenqtscase2_3}
	$\tilde{\wp}'_\text{cs}(n;\alpha,\beta)$
	for $\alpha=1$ and~$\beta=6$.%
}
\label{fig:pndevenqtscase2}
\end{figure}

Figure~\ref{subfig:pndevenqtscase2_1} shows the photon-number distribution for this arrangement of the QTS~$\Upsilon2$ . The two Poissonians are peaked at 1 and 36 for the QDS $\ket{\pm1}$ and QDS $\ket{\pm 6}$, respectively. As the separation between the QDS formed by~$\ket{\pm1}$ is small, the Poissonian curve peaked at $1$ is narrow. Figure \ref{subfig:pndevenqtscase2_2} shows the interference effect as in Eq.~(\ref{eq:interPoissonianinterference}) plotted against $n$. It shows the interference arising between the two QDSs $\ket{\pm1}$ and $\ket{\pm6}$ is the smallest among all the cases probably because the pair~$\ket{\pm1}$ does not have enough separation between them.
As expected, Fig.~\ref{subfig:pndevenqtscase2_3} shows that the small inter-Poissonian interference does not change the photon-number distribution significantly at the envelope level.
\subsubsection{$\Upsilon2$: Quadruple-Gaussian-well ground state}
\begin{figure}
\centering
\begin{subfigure}{0.40\textwidth}
\includegraphics[width=\linewidth]{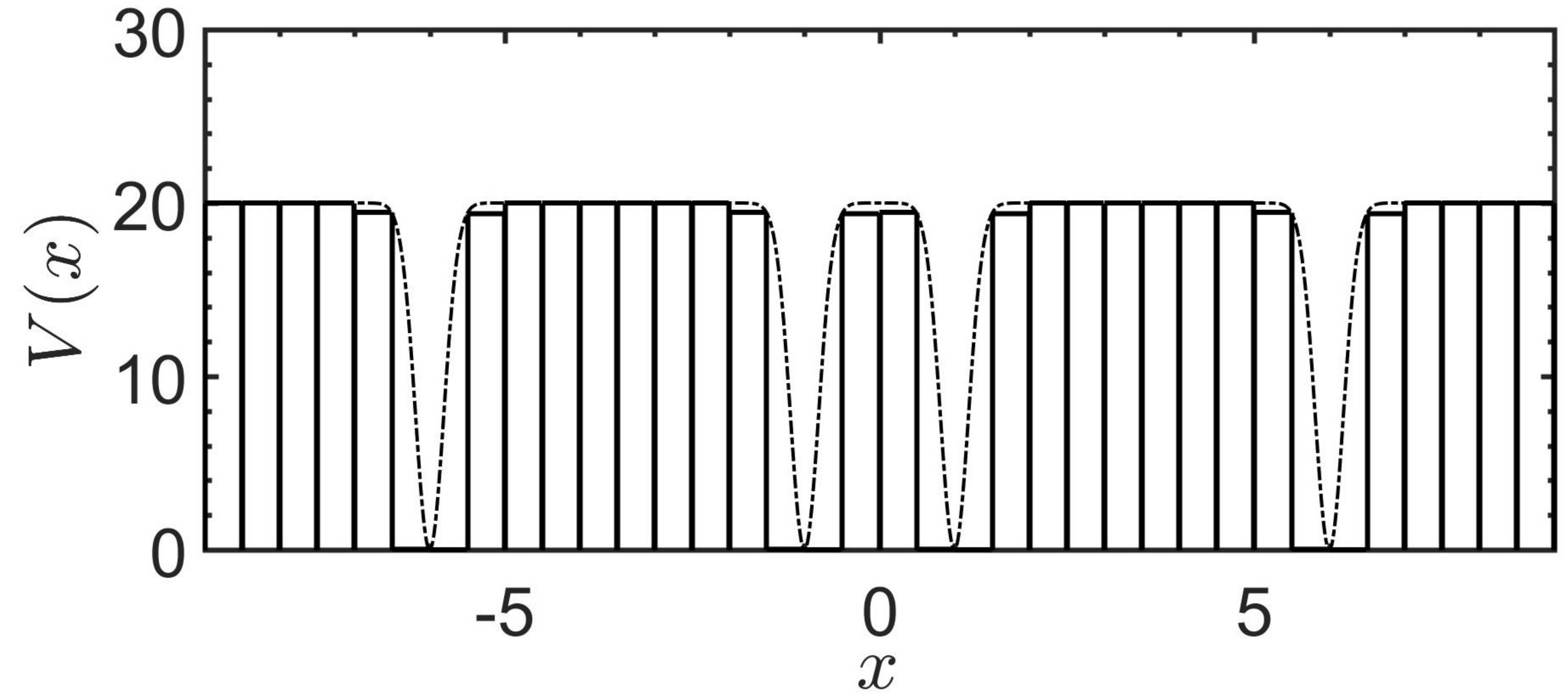}
\caption{ }
\label{subfig:gaussianwellscase2_1}
\end{subfigure}
\vspace*{-0.1cm}
	\hspace*{3cm}
  \flushleft
\begin{subfigure}{0.44\textwidth}
\hspace*{0.5cm}
\includegraphics[width=\linewidth]{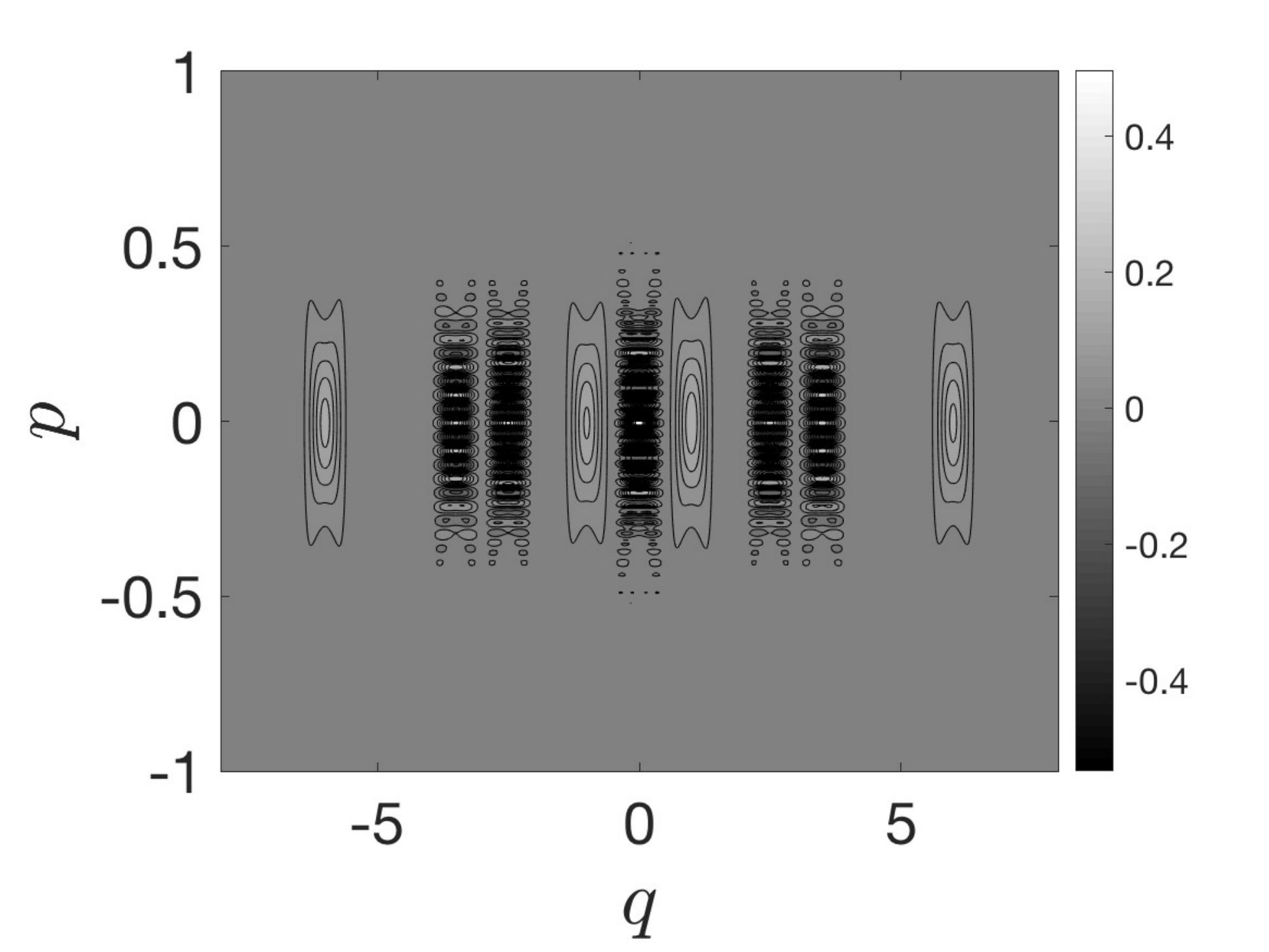}
\caption{ }
\label{subfig:gaussianwellscase2_2}
\end{subfigure}
\caption{$\Upsilon2$:~\subref{subfig:gaussianwellscase2_1} quadrupole-Gaussian-well potential approximated to piecewise continuous function and~\subref{subfig:gaussianwellscase2_2} ground-state Wigner function for the potential $V(x)$ for $\alpha=1,\beta=6$ ($\Upsilon2$).}
\label{fig:gaussianwellscase2}
\end{figure}

The four-well Gaussian potential for this case is depicted in Fig.~\ref {subfig:gaussianwellscase2_1}. The Wigner function calculated for the approximated ground state of this potential is shown in Fig.~\ref{subfig:gaussianwellscase2_2}. The locations of Gaussian peaks and the centres of the interference patterns between each pair of coherent states agree with Eq.~(\ref{eq:WignerQTS})
and plotted in Fig.~\ref{subfig:Wignermarginal2_1}. 
\subsection{$\Upsilon3$: Comb State}
\label{subsec:Upsilon3}
In this subsection,
we analyze the comb-state QTS ($\Upsilon3$).
Specifically,
we analyze the results for the Wigner function,
the marginal distributions,
the photon-number distribution,
and the four-well ground-state approximation.
\subsubsection{$\Upsilon3$: Wigner function}
\begin{figure}
\centering
\begin{subfigure}{0.40\textwidth}
\includegraphics[width=\linewidth]{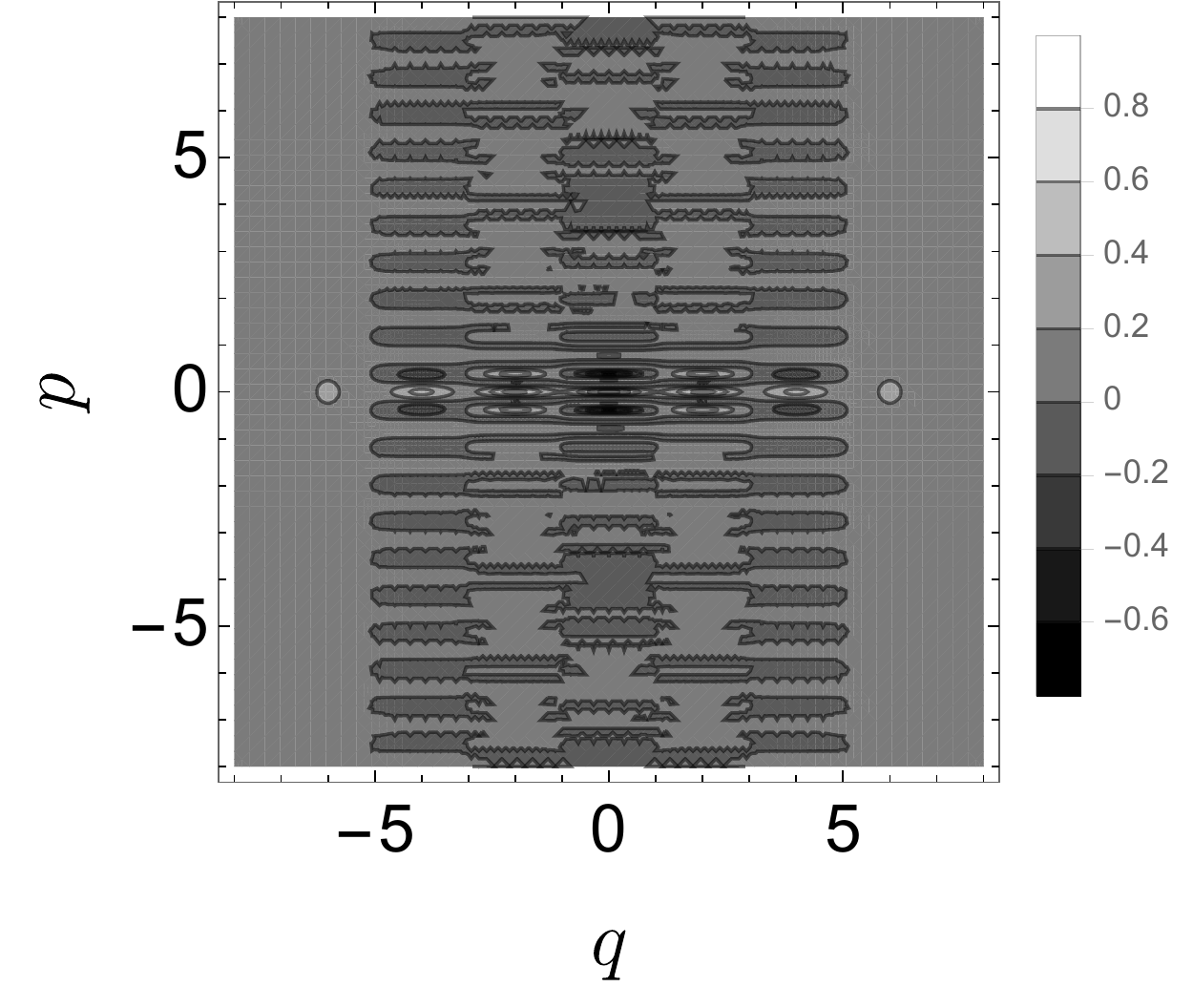} 
\caption{ }
\label{subfig:Wignermarginal3_1}\vspace{0.5 cm}
\end{subfigure}
\begin{subfigure}{0.23\textwidth}
\includegraphics[width=\linewidth]{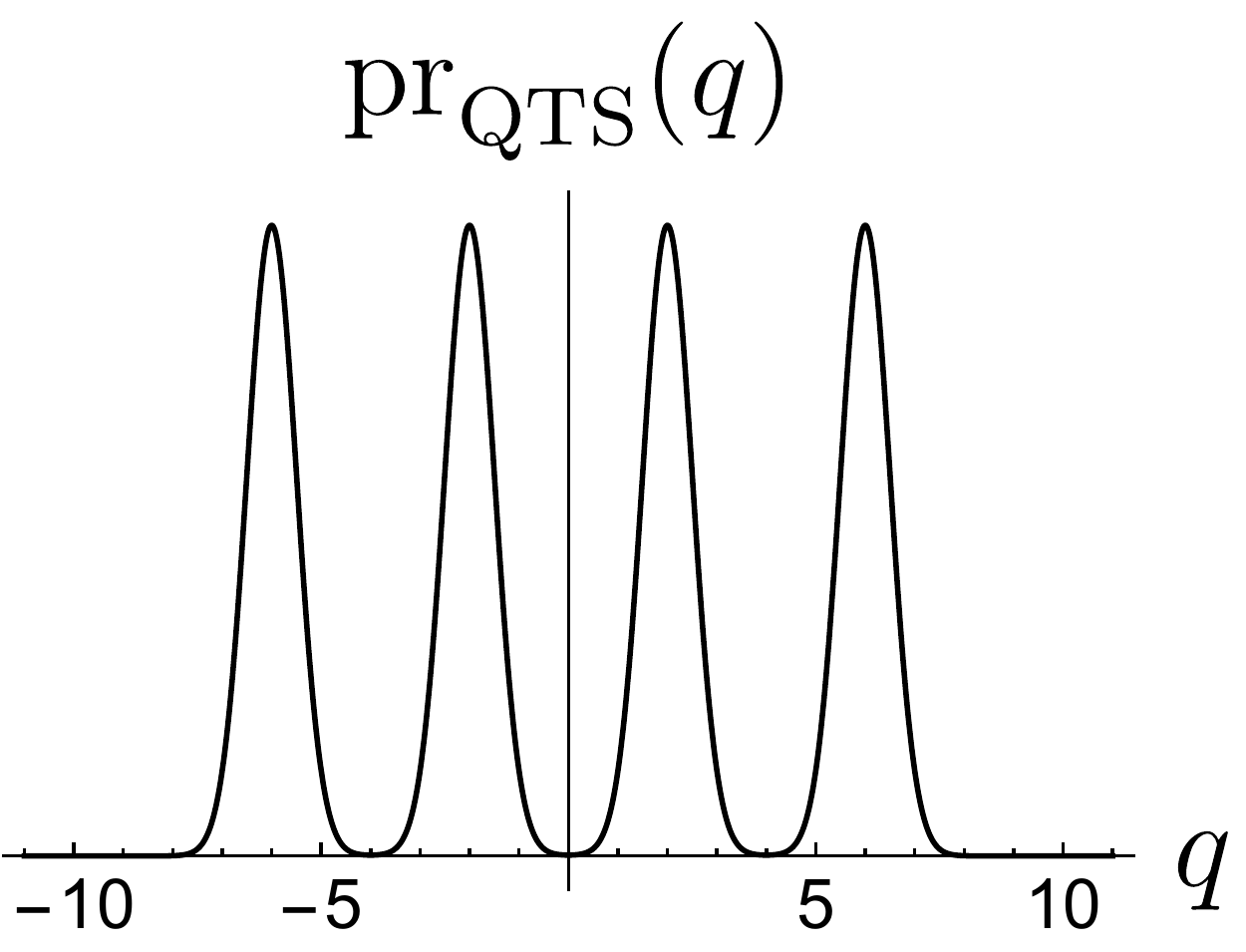}
\caption{ }
\label{subfig:Wignermarginal3_2}
\end{subfigure}
\hspace{0.1 cm}
\begin{subfigure}{0.23\textwidth}
\includegraphics[width=\linewidth]{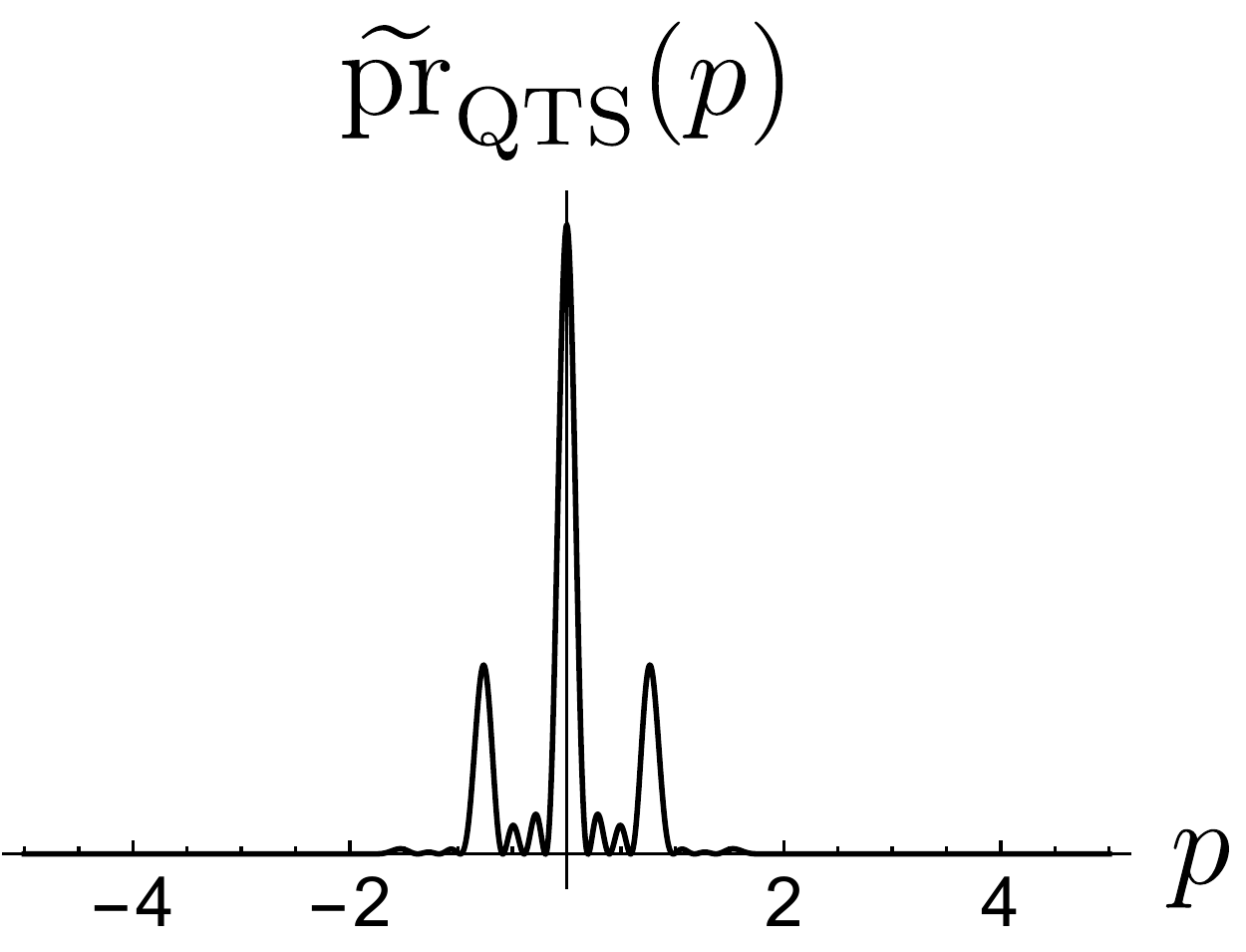}
\caption{ }
\label{subfig:Wignermarginal3_3}
\end{subfigure} 
\caption{Wigner function and marginal distributions
	for the QTS case~$\Upsilon3$ for  $\alpha=2,\beta=6$.}
\label{fig:Wignermarginal3}
\end{figure}

The third case as shown in Fig.~\ref{subfig:phasespace picture3} corresponds to a cat having two states each split into doublets with the same overlap as the two doublets ($\Upsilon3$).
The Wigner function for this case is depicted in Fig.~\ref{subfig:Wignermarginal3_1} and it shows only one pair of Gaussians corresponding to~$\ket{\pm6}$ centred at $\pm6$ on the position axis.
Whenever $3\alpha=\beta$,
which is the case here,
interference coincides with Gaussians peaks resulting a spreading of the interference pattern along the $q$ axis. 
\subsubsection{$\Upsilon3$: Marginal distributions}
The QTS marginal distributions for configuration~$\Upsilon2$ are shown in Figs.~\ref{subfig:Wignermarginal3_2} and~\ref{subfig:Wignermarginal3_3}.
We see the Gaussian peaks at $\pm2$ and $\pm6$ along the position axis and an interference Gaussian modulated by $(\cos2p+\cos6p)^2$ along the momentum quadrature. 

\subsubsection{$\Upsilon3$: Photon-number distribution}
\begin{figure}
\centering
\begin{subfigure}{0.40\textwidth}
\includegraphics[width=\linewidth]{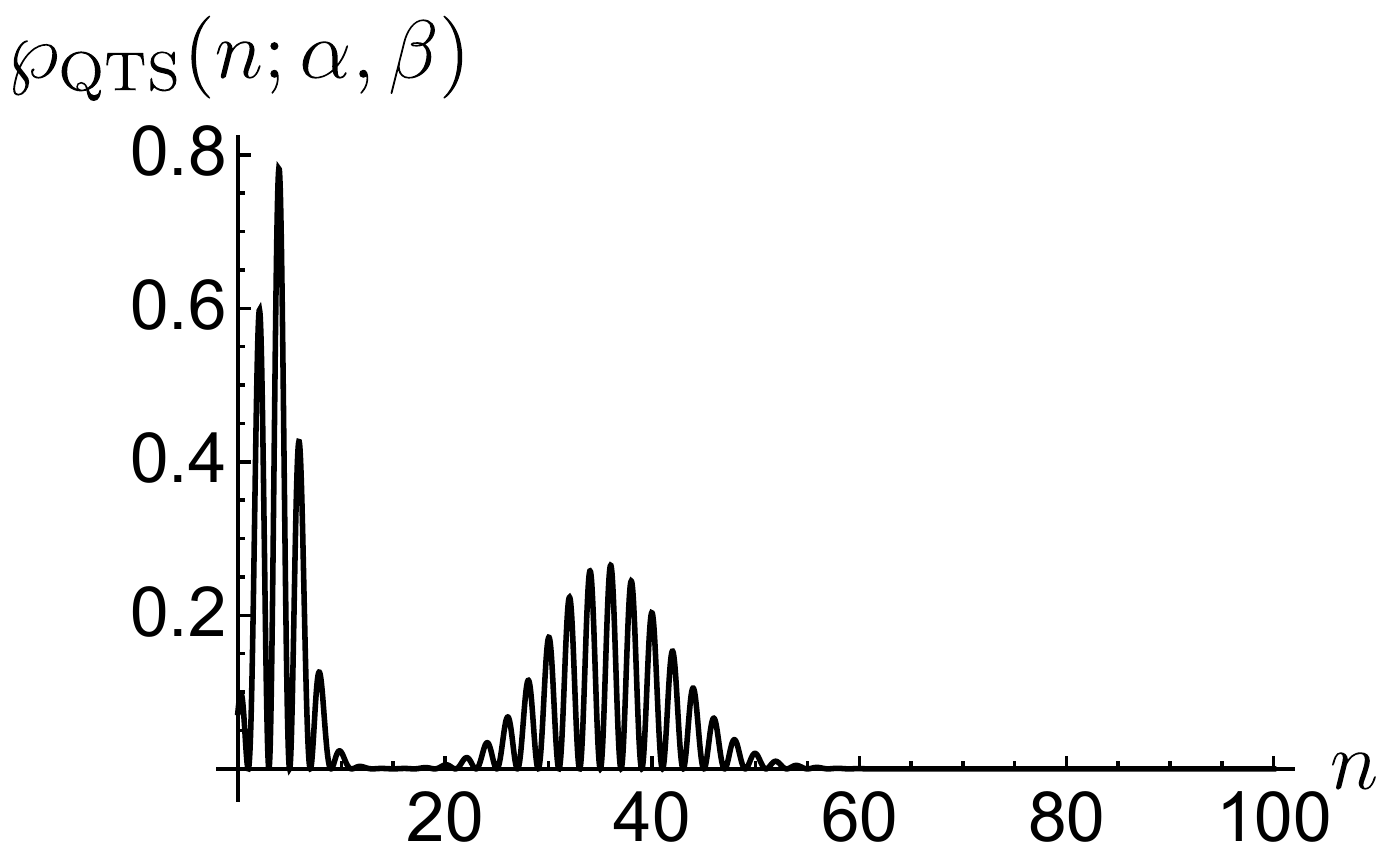} 
\caption{ }
\label{subfig:pndevenqtscase3_1}
\vspace{0.5 cm}
\end{subfigure}
\begin{subfigure}{0.23\textwidth}
\includegraphics[width=\linewidth]{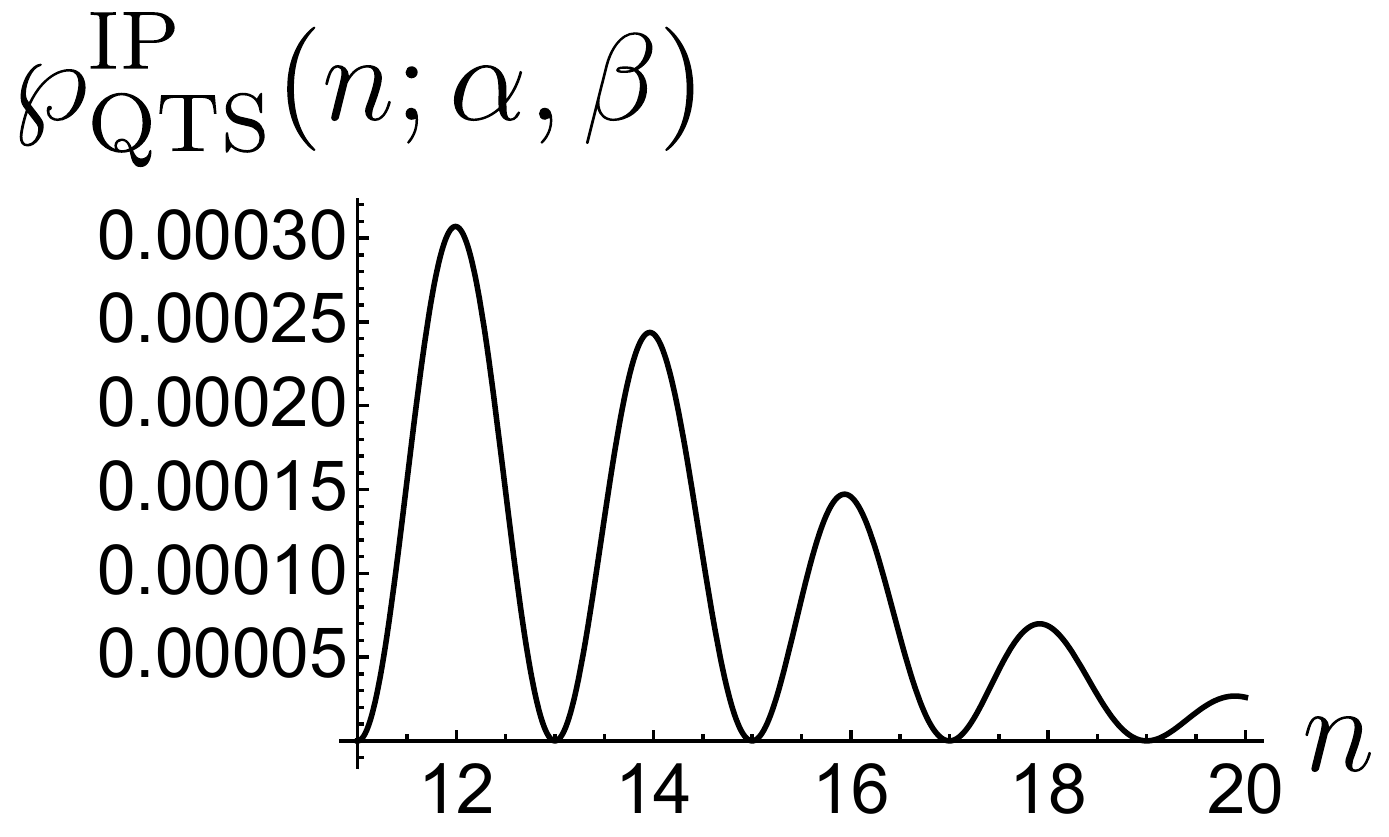}
\caption{ }
\label{subfig:pndevenqtscase3_2}
\end{subfigure}
\hspace{-0.1 cm}
\begin{subfigure}{0.23\textwidth}
\includegraphics[width=\linewidth]{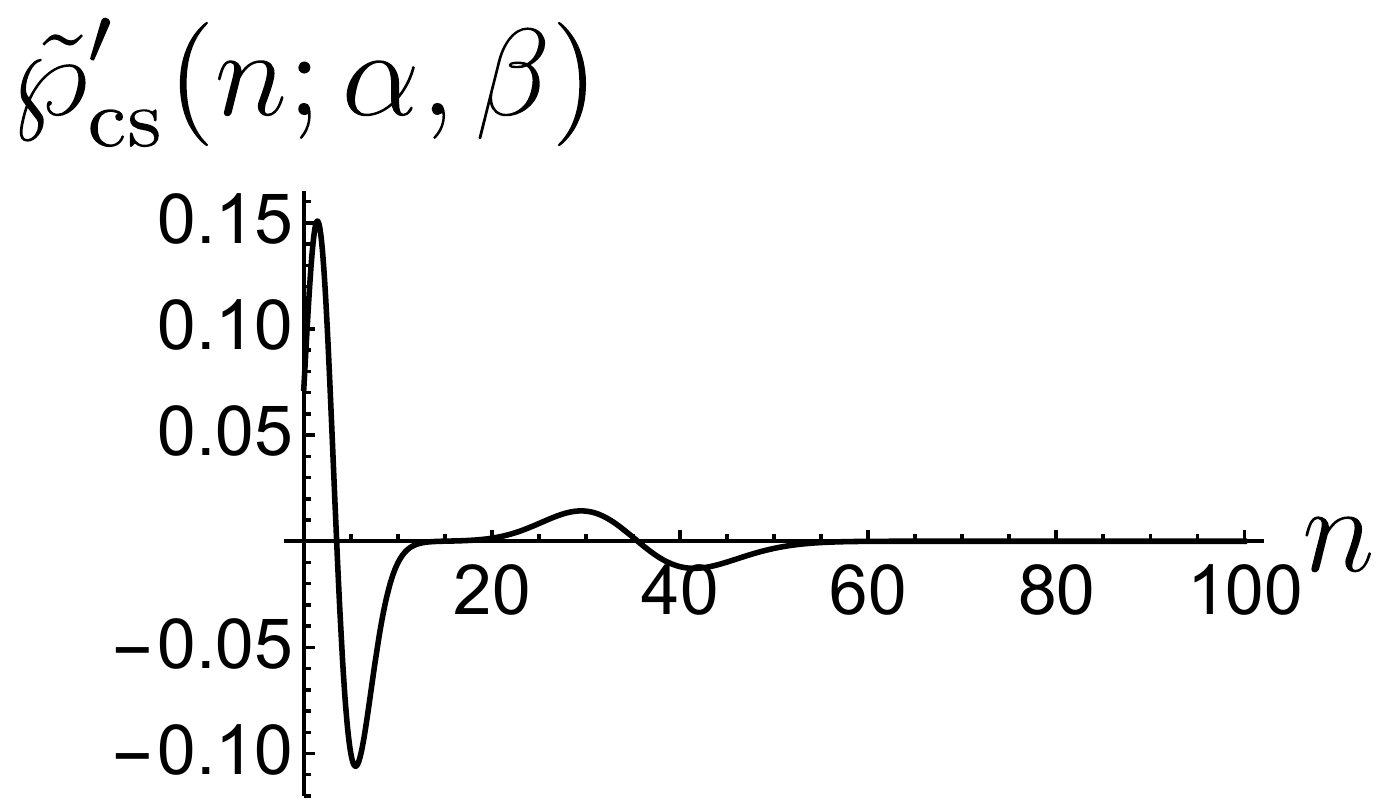}
\caption{ }
\label{subfig:pndevenqtscase3_3}
\end{subfigure}
\caption{%
	$\Upsilon3$:
	Plots of
	\subref{subfig:pndevenqtscase3_1}
	$\wp_\text{QTS}(n;\alpha,\beta)$,
	\subref{subfig:pndevenqtscase3_2}
	$\wp^\text{IP}_\text{QTS}(n;\alpha,\beta)$
	and
	\subref{subfig:pndevenqtscase3_3}
	$\tilde{\wp}'_\text{cs}(n;\alpha,\beta)$
	for $\alpha=2$ and~$\beta=6$.%
}
\label{fig:pndevenqtscase3}
\end{figure}

The photon-number distribution in this QTS arrangement for configuration~$\Upsilon2$ appears in Fig.~\ref{subfig:pndevenqtscase3_1}. 
Poisson distributions are clearly well separated with their highest peaks located at $4$ and $36$ in phase space because the QDSs $\ket{\pm2}$ and $\ket{\pm6}$ are well separated.
A closer view of inter-Poissonian interference~(\ref{eq:interPoissonianinterference})
is depicted in Fig.~\ref{subfig:pndevenqtscase3_2}.
Comparing to previous cases,
interference in this case is more pronounced than for the QTS configuration for~$\Upsilon2$ but less than for~$\Upsilon1$.
The derivative of the envelope curve without the inter-Poissonian interference term (\ref{eq:pd'QTS_2}) in Fig.~\ref{subfig:pndevenqtscase3_3}
shows that the interference between the Poissonian distributions  does not affect the maxima and minima of the photon-number distribution.
\subsubsection{$\Upsilon3$: Quadruple-Gaussian-well ground state}
\begin{figure}
\centering
\begin{subfigure}{0.40\textwidth}
\includegraphics[width=\linewidth]{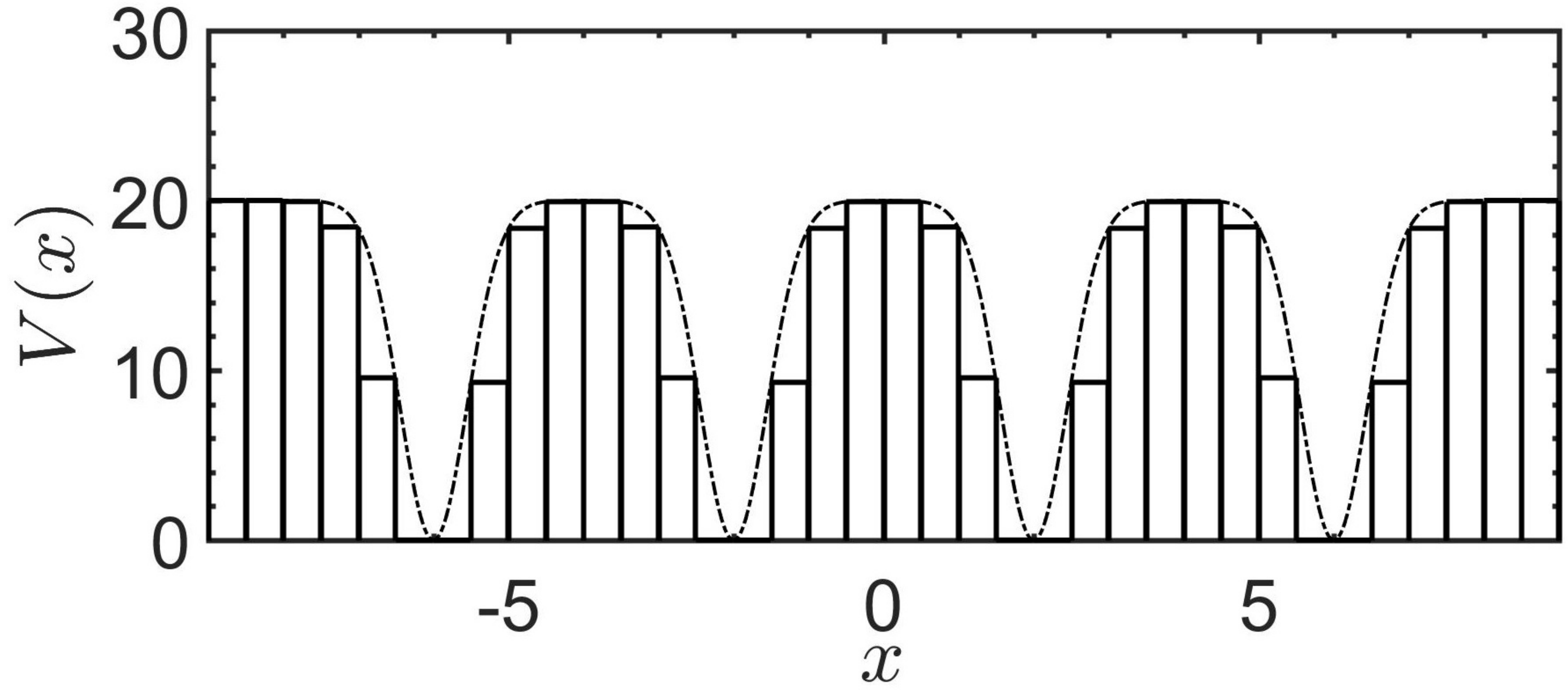}
\caption{ }
\label{subfig:gaussianwellscase3_1}
\end{subfigure}
\vspace*{-0.1cm}
	\hspace*{3cm}
  \flushleft
\begin{subfigure}{0.40\textwidth}
\hspace*{0.5cm}
\includegraphics[width=\linewidth]{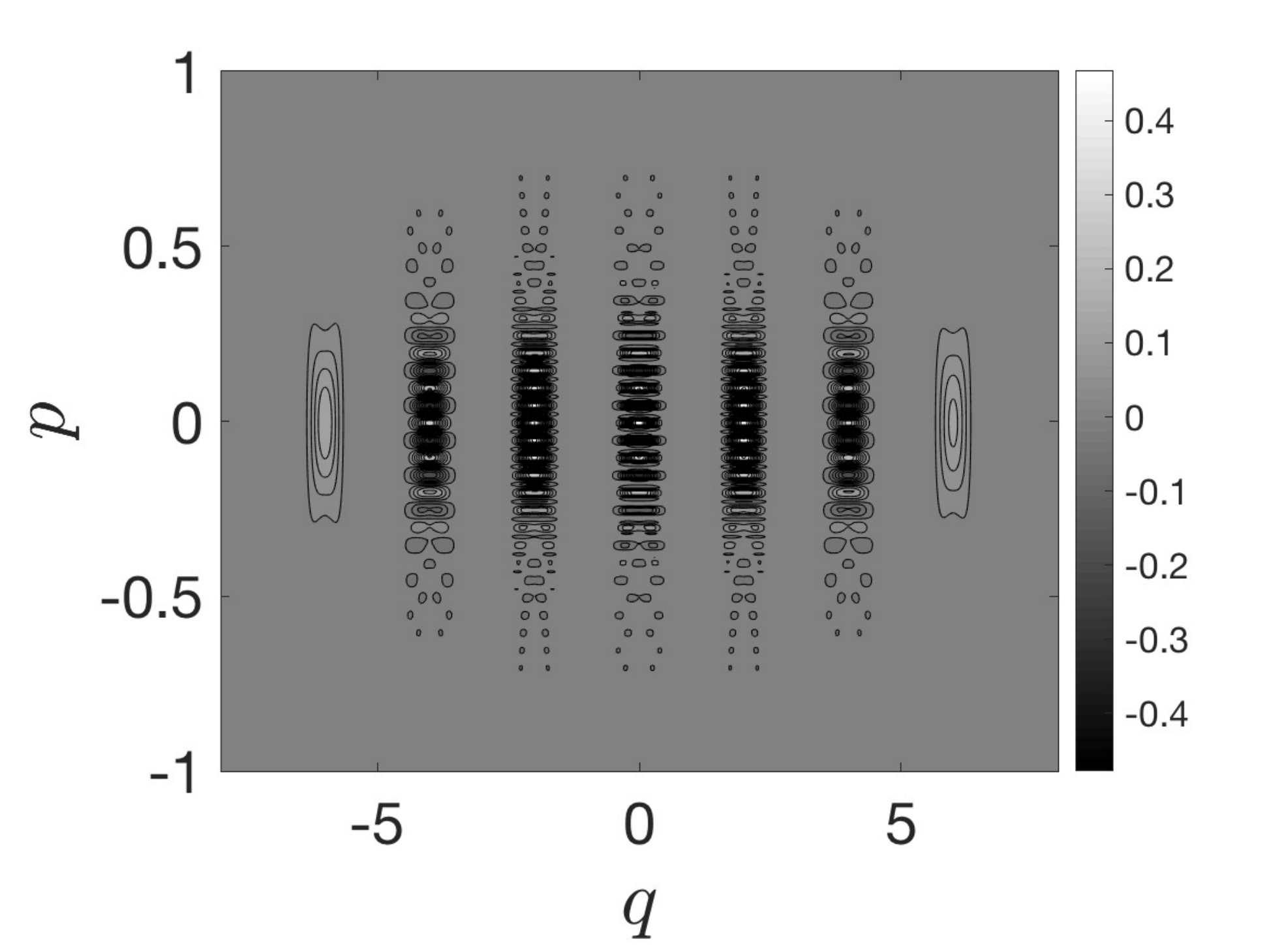}
\caption{ }
\label{subfig:gaussianwellscase3_2}
\end{subfigure}
\caption{$\Upsilon3$:~\subref{subfig:gaussianwellscase3_1} quadrupole-Gaussian-well potential approximated to piecewise continuous function and~\subref{subfig:gaussianwellscase3_2} ground-state Wigner function for the potential $V(x)$ for $\alpha=2,\beta=6$ ($\Upsilon3$).}
\label{fig:gaussianwellscase3}
\end{figure}

The Wigner function calculated for the approximate ground-state of quadruple-Gaussian-well potential arrangement shown in Fig.~\ref{subfig:gaussianwellscase3_1} is depicted in Fig.~\ref{subfig:gaussianwellscase3_2}.
The Wigner-function interference patterns for this approximated ground state coincide with the Wigner function that has been analytically constructed for this QTS in Fig.~\ref{subfig:Wignermarginal3_1}.
\section{Discussion}
\label{sec:discussion}
In this section, we interpret and summarize the results for each of the three cases discussed for the four tools comprising the Wigner function,
the marginal distributions
and the photon-number distribution.
One intuitive result is that the Wigner function for each QTS is understood,
due to linearity,
as a superposition of the $\binom{4}{2}=6$ QDS phenomena.

This insight helps to see clearly what the otherwise complicated Wigner-function patterns are showing.
Interference in the Wigner function is quite evident if the QDSs are greatly separated between coherent-state components
(the cat-state case)
and exhibit interference poorly for small separation between coherent states
(the kitten-state case).
If interference coincides with  a Gaussian peak,
the Gaussian peak is thus diminished, which is important for the comb states, which has equal spacing so most Gaussians are co-located with strong interference so the comb state displays almost vanishing Gaussians
(the two in the middle of the four-coherent-state case).

For marginal distributions,
the Gaussian peaks are well separated and interference along the $p$-axis is clear.
For the kitten state, 
Gaussian peaks run together.
The $p$-distributions also show fading interference as distance between coherent states decreases.

The PND for the QTS is a an oscillatory function that is modulated by two Poissonian peaks. This modulation is primarily a QDS effect explained by interference in phase space, but the oscillation between peaks is a QTS effect.
The minima and maxima of the PND envelope do not shift as evidenced by the 
expression for the derivative of the PND.

We have discussed a physical implementation of each QTS by obtaining the ground state of a quadruple-Gaussian-well potential.
Our numerical analysis shows excellent agreement between the QTS and the ground state
with respect to where the Gaussian peaks and interference effects appear in phase space and how they look.
Some differences arise due to the Gaussian well not corresponding perfectly to parabolic wells.
\section{Conclusions}
\label{sec:conclusions}
To conclude, the QTS is introduced to extend the idea of so-called Schr\"{o}dinger cat states
by effectively splitting each of the two coherent states in the superposition into two,
thereby obtaining the QTS as a superposition of four coherent states.
We have calculated and plotted the Wigner functions, marginal distributions and photon-number distributions
and explained the salient features,
which appear complicated but can be understood in terms of linear superpositions and beat patterns.

Different cases were presented corresponding to a macroscopically separated superposition of Schr\"{o}dinger kitten states,
a microscopically separated superposition of Schr\"{o}dinger cat states
and a four-tooth version of the quantum comb state (in the infinite limit)~\cite{GKP01}.
The features have been explained,
and the comb-state analysis could be especially useful as these features have not been explored previously yet are meaningful in experimental efforts towards quantum information processing in a harmonic oscillator.

Another important aspect of our work is the connection between the QTS and the quadruple-well potential.
We solve for Gaussian wells but of course other wells are possible.
Multiple potential wells provide a promising avenue for guiding laboratory realizations of the QTS.
\acknowledgments
N.S.\ acknowledges the University of Calgary Eyes High Postdoctoral program.
N.A.\ acknowledges the China Scholarship Council (Grant No.\ 2016GXYN31).
B.C.S.\ acknowledges NSERC and NSFC (Grant No.\ 11675164) for financial support.
\bibliography{tetrachotomous}
\end{document}